\documentclass[12pt]{article}

\usepackage{youngtab}
\usepackage{a4}
\usepackage{amsfonts}
\usepackage{amssymb}
\usepackage{lscape}
\usepackage{cite}
\usepackage{epsfig}
\usepackage{multirow}
\usepackage[all]{xy}
\usepackage{amsmath}
\usepackage{amssymb}
\let\a=\alpha   \let\b=\beta

\def\slash#1{\mbox{$\not \!\! #1$}}

\newcommand{\beq}{\begin{equation}}
\newcommand{\eeq}{\end{equation}}
\newcommand{\beqn}{\begin{eqnarray}}
\newcommand{\eeqn}{\end{eqnarray}}

\newcommand{\nn}{\nonumber}

\newcommand{\mP}{\mathbb{P}}
\newcommand{\mZ}{\mathbb{Z}}
\newcommand{\mC}{\mathbb{C}}
\newcommand{\mR}{\mathbb{R}}

\newcommand{\be}{\begin{equation}}
\newcommand{\ee}{\end{equation}}
\newcommand{\ba}{\begin{eqnarray}}
\newcommand{\ea}{\end{eqnarray}}
\newcommand{\bdm}{\begin{displaymath}}
\newcommand{\edm}{\end{displaymath}}

\def\b{\beta}

\def\a{\alpha}

\newcommand{\ie}{{\it i.e.\ }}
\newcommand{\eg}{{\it e.g.\ }}

\DeclareMathAlphabet{\mathpzc}{OT1}{pzc}{m}{it}
\def\bea{\begin{eqnarray}}
\def\eea{\end{eqnarray}}
\def\beas{\begin{eqnarray*}}
\def\eeas{\end{eqnarray*}}
\def\sla{\raise.15ex\hbox{$/$}\kern-.57em}

\def\bea{\begin{eqnarray}}
\def\eea{\end{eqnarray}}
\def\de{\partial}

\def\sla{\raise.15ex\hbox{$/$}\kern-.57em}
\def\ie{{\it i.e.}~}
\def\eg{{\it e.g.}~}
\def\ap{{\alpha^\prime}}

\def\a{\alpha}
\def\b{\beta}

\def\cA{{\cal A}}

\def\cD{{\cal D}}
\def\cE{{\cal E}}
\def\cF{{\cal F}}

\def\cH{{\cal H}}

\def\cJ{{\cal J}}

\def\cM{{\cal M}}
\def\cN{{\cal N}}

\def\cR{{\cal R}}

\def\cX{{\cal X}}

\def\cZ{{\cal Z}}

\begin{document}
\begin{titlepage}
\begin{flushright}
{ROM2F/2010/09}\\
\end{flushright}
%
%
\begin{center}
{\Large\bf Precision Spectroscopy and Higher Spin symmetry in the ABJM model}  \\
\end{center}
\begin{center}
{\bf Massimo Bianchi$^{1,2}$, Rubik Poghossian$^{1,3}$ and Marine Samsonyan$^1$}\\
{\sl $^1$Dipartimento di Fisica, Universit\`a di Roma ``Tor Vergata''\\
 I.N.F.N. Sezione di Roma ``Tor Vergata''\\
Via della Ricerca Scientifica, 00133 Roma, Italy}\\
and \\
 {\sl $^2$Physics Department, Theory Unit, CERN \\ CH 1211,
Geneva 23, Switzerland }\\
and\\
{\sl $^3$Yerevan Physics Institute\\
Alikhanian Br. 2, 0036 Yerevan, Armenia}
\end{center}
\vskip 1.0cm
\begin{center}
{\large \bf Abstract}
\end{center}
We revisit Kaluza-Klein compactification of 11-d supergravity on
$S^7/\mZ _k$ using group theory techniques that may find application
in other flux vacua with internal coset spaces. Among the $SO(2)$
neutral states, we identify marginal deformations and fields that
couple to the recently discussed world-sheet instanton of Type IIA
on $\mC \mP^3$. We also discuss charged states, dual to monopole
operators, and the $\mZ _k$ projection of the $Osp(4|8)$ singleton
and its tensor products. In particular, we show that the doubleton
spectrum may account for $\cN =6$ higher spin symmetry enhancement
in the limit of vanishing 't Hooft coupling in the boundary
Chern-Simons theory.



\vfill

\end{titlepage}

\section*{Introduction}

The spectrum of Kaluza-Klein (KK) excitations in flux vacua plays
an important role both in attempts to embed the Standard Model in
String Theory and in the holographic correspondence. In the spirit
of holography, the seminal observation of Schwarz's
\cite{Schwarz:2004yj} and the subsequent work of Bagger and
Lambert \cite{Bagger:2006sk, Bagger:2007jr, Bagger:2007vi,
Bagger:2008se } and, independently, of Gustavsson
\cite{Gustavsson:2007vu }, motivated Aharony, Bergman, Jafferis
and Maldacena (ABJM) \cite{Aharony:2008ug,Klebanov:2009sg} to propose a duality between
superconformal Chern-Simons (C-S) theories in $d=3$ dimensions and String /
M- theory on $AdS_4$.

The duality has been thoroughly tested and extended to cases with
lower supersymmetry \cite{Aganagic:2009zk, Jafferis:2009th, Gaiotto:2009tk, Jafferis:2008em,
Jafferis:2008qz, Forcella:2009jj}. In particular the
superconformal index has been matched both in the regime $k>>1$
($SO(2)$ singlets) \cite{Bhattacharya:2008bja,
Bhattacharya:2008zy} and at finite $k$ \cite{Kim:2009wb,Kim:2010vw
}. A detailed analysis of the (BPS) spectrum and the
supermultiplet structure is however still incomplete. Aim of this
note is to fill in this gap and perform precision spectroscopy of
11-d supergravity on $AdS_4\times S^7/\mZ_k$ or, equivalently, Type
IIA on $AdS_4\times \mC\mP^3$. We will also discuss higher spin
symmetry enhancement in the limit of vanishing 't Hooft coupling
in the boundary $\cN =6$ Chern-Simons theory.

The plan of the paper is as follows. After reviewing the ABJM
model, presenting both bulk and boundary vantage points, we will
revisit KK reduction of 11-d supergravity on $S^7$
\cite{Castellani:1984vv} and then perform the decomposition of
$SO(8)$ into $SO(6)\times SO(2)$ so as to derive the KK
excitations of $\cN=6$ gauged supergravity \cite{Salam:1989fm},
including states charged under $SO(2)$ that are expected to be
dual to `monopole' operators on the boundary
\cite{Aharony:2008ug,Klebanov:2009sg}. Since we rely on group theory
techniques which are not easily found in the available literature,
we try to make this part of the presentation as pedagogical as
possible, also in view of applications to other flux vacua with
internal coset manifolds $G/H$. We then compare the resulting bulk
spectrum with the spectrum of gauge-invariant operators on the
boundary. Finally we compute the partition function of the
boundary theory performing an orbifold projection on the parent
theory ($k=1,2$ cases) and examine the higher spin content of the
theory. Various appendices summarize useful $SO(8)$ and $SO(6)$
group theory formulae.

\section{The ABJM model}

The near-horizon geometry of a stack of $N$ M2-branes is
$AdS_4\times S^7$ with $N$ units of $F_4$ flux along $AdS_4$ and
as many units of its dual $F_7$ along $S^7$ \cite{Malda}. The
metric reads \be ds^2_{11} = {1\over 4} L^2 ds^2_{AdS} + L^2
ds^2_{S^7} \ee for later use, note that $L_{AdS} = L/2$ with $L$
the radius of $S^7$ and henceforth the metrics of the subspaces are
for unit curvature radii.

ABJM have conjectured that 11-d supergravity on $AdS_4\times
S^7/\mZ_k$, corresponding to the near horizon geometry of $N$
M2-branes at a $\mC^4/\mZ_k$ singularity, be dual to $\cN=6$ C-S
theory in $d=3$ with gauge group $U(N)_k\times U(N)_{-k}$ and
opposite CS couplings $k_1=k=-k_2$ \cite{Aharony:2008ug, Klebanov:2009sg}.

\subsection{Supergravity description}

The Type IIA solution corresponding to the ABJM model reads \be
ds^2_{IIA} = 4 {\rho^2\over L^2} dx\cdot dx + 4 {L^2\over 4
\rho^2} d\rho^2 + L^2 ds^2_{\mC\mP^3} = {1\over 4} L^2 ds^2_{AdS} +
L^2 ds^2_{\mC\mP^3} \ee where \be L = \left(32\pi^2 N \over k
\right)^{1/4} \ee is the curvature radius in string units. The
string coupling, related to the VEV of the dilaton, is given by
\be g_s = L/k = \left(32\pi^2 N \over k^5 \right)^{1/4} \ee Thus
the perturbative Type IIA description should be valid for $L>>1$
and $g_s<<1$ \ie for $N^{1/5}<<k<<N$ while $\lambda=N/k$ is the 't
Hooft coupling of the boundary CS theory.

In the 11-d uplift, ${\mC\mP^3}$ becomes the base of a Hopf fibration
$S^7 = \mC\mP^3 \ltimes S^1$ whose metric reads \be ds^2_{S^7} =
ds^2_{\mC\mP^3} + (d\tau + \cA)^2 \ee with $d\cA = 2 \cJ_{\mC\mP^3}$, the
K\"ahler form on $\mC\mP^3$ normalized so that $dV(\mC\mP^3)=\cJ\wedge \cJ
\wedge \cJ/6$ and $V(\mC\mP^3) = \pi^3/6$. The solution is supported
by R-R fluxes \be g_s F_2= 2L \cJ \quad , \quad g_s F_4= {6 L^3}
dV(AdS_4) \quad , \quad g_s F_6= 6 L^5 dV(\mC\mP^3)\ee

In the ABJM model, corresponding to $\cN = 6$ C-S theory $U(N)_{k}
\times U(N)_{-k}$ on the boundary, $B_2 = 0$. For fractional
M2-branes, one has the ABJ model corresponding to $\cN = 6$ C-S
theory $U(N)_{k} \times U(N+k-l)_{-k}$ \cite{Aharony:2008gk} on
the boundary, $B_2 = \cJ l/k$, with $l=1, ..., k-1$. Boundary C-S
theories with $\sum_i k_i \neq 0$ and lower susy should be dual to
turning on a non-zero Romans mass ($F_0\neq 0$) in the bulk Type
IIA description \cite{Gaiotto:2009mv, Petrini:2009ur, Gaiotto:2009yz}.

The 11-d supergravity approximation should be valid in the
double-scaling limit $k\rightarrow \infty$, $N\rightarrow \infty$
with $\lambda = N/k$ fixed and large. The CFT description, to
which we momentarily turn our attention, should instead be valid
when $\lambda<<1 $, \ie $k>>N$. As $\lambda \rightarrow 0$ higher
spin symmetry enhancement takes place as we will eventually see.

\subsection{Boundary CFT description}

$\cN =6$ CS theories are conveniently constructed from $\cN =3$ CS
theories. The case $\cN =3$ arises in turn from the $\cN =4$ case
obtained after dimensional reduction of $\cN' = 2$ in $d=4$. In
this way, each vector multiplet includes an $\cN = 2$ (\ie
$\cN=1'$ in $d=4$) chiral multiplet in the adjoint $\Phi = \Phi_a
t^a$ and couples to various hypers $Q$ and $\tilde{Q}$ in real
(reducible) representations. Adding to the `standard' $\cN = 4$
superpotential \be W = \tilde{Q} \Phi Q \ee the CS term, giving a
mass $m = g_{YM}^2 {k\over 4\pi}$ to the vectors, and a CS
superpotential \be W = {k\over 8\pi} Tr\Phi^2 \ee breaks $\cN =4$
to $\cN =3$. Integrating out $\Phi$ yields \be W= {4\pi \over k}
(\tilde{Q} t^a Q) (\tilde{Q} t^a Q) \ee The resulting $\cN =3$
theory has no marginal susy preserving deformations
\cite{Gaiotto:2009yz, Petrini:2009ur, Gaiotto:2009mv}. In
the process R-symmetry is reduced to $SO(3)\approx SU(2)$ for $\cN
=3$ from the original $SO(4)$ of $\cN =4$.

The case $\cN = 6$ is special. Starting with the $\cN =3$ theory
with $G=U(N)_k \times U(N)_{-k}$ and two pairs of hypers, $A_r \in
({\bf N}, {\bf N}^*)$ and $B_{\dot{m}} \in ({\bf N}^*, {\bf N})$
and integrating out $\Phi_1$ and $\Phi_2$ one gets \be W = {2\pi
\over k} \epsilon^{rs} \epsilon^{\dot{m} \dot{n}} Tr(A_r
B_{\dot{m}} A_s B_{\dot{n}}) \ee Since the manifest `flavour'
symmetry of $W$ under $SU(2)\times SU(2) \times U(1)_B$ does not
commute with R-symmetry $SO(3)\approx SU(2)$ under which $A$ and
$B$ form doublets, the full theory has a larger $SU(4)\approx
SO(6)$ symmetry which is the R-symmetry of $\cN=6$. To expose the
symmetry it is convenient to define $X^i = (A_1, A_2,
B^*_{\dot{1}}, B^*_{\dot{2}})$ and their conjugate $X_i^*$ that
together transform as ${\bf 4}_{+1} + {\bf 4}^*_{-1}$ of
$SO(6)\times SO(2)$. As we will momentarily see, $SO(2)\sim U(1)$
acts as a baryonic symmetry. Further (super)symmetry enhancement
to $\cN = 8$ with $SO(8)$ R-symmetry takes place for $k=1$ and
$k=2$. The former corresponds to compactification on $S^7$ the
latter to $S^7/Z_2$ (only `even' spherical harmonics).

\subsection{A quick look at the spectrum}

The (ungauged) $\cN = 6$ supergravity multiplet consists of the
graviton $g_{\mu\nu}$, 6 gravitini $\psi_\mu^i$, 16 graviphotons
$A_\mu^{[ij]}$ and $A_\mu^{0}$, 26 dilatini $\lambda^{[ijk]}$ and
$\lambda_i$, and 30 scalars $\phi^{[ijkl]}$ and $\phi_{[ij]}$. The
latter parameterize the moduli space $\cM = SO^*(12)/U(6)$.  After
`gauging' $SO(6)\times SO(2)$ a scalar potential is generated and
the two sets of ${\bf 15}_0$ scalars become `massive' or rather
`tachyonic' \ie $(ML_{AdS})^2 = -2$, safely above the B-F bound
$(ML_{AdS})^2=-9/4$.

Compactification of Type IIA supergravity on $\mC\mP^3$ was studied in
\cite{Nilsson:1984bj}. KK excitations with $Q=0$, \ie neutral wrt
$SO(2)$, were identified there. The non-perturbative spectrum,
contains various wrapped branes, including D0-branes that are
charged wrt $SO(2)$. The latter correspond to 11-d KK modes along
the compact circle that can be obtained by a $\mZ_k$ projection of
the M-theory compactification on $S^7$. The dual to $SO(2)$
charged states are monopole operators on the boundary
\cite{Klebanov:2008vq, Benna:2009xd, Klebanov:2009sg}. Although
the fundamental fields ($A_r, B_{\dot{s}}$) are neutral wrt the
diagonal $U(1)$ that couples to $A_\mu^+ = A_\mu^1 + A_\mu^2$, the
orthogonal combination $A_\mu^- = A_\mu^1 - A_\mu^2$ acts as a
baryonic symmetry. The corresponding current, $J_B = *F^+$, is
conserved thanks to Bianchi identities. Due to the CS coupling $
k\int A^-\wedge F^+$, configurations with $A^+$ magnetic charge
are electrically charged wrt $A^-$. Alternatively one can
introduce a Lagrange multiplier $\tau$ for $dF^+=0$ (on-shell
$kA^-= d\tau$) and form combinations $e^{in\tau}$ that can screen
the baryonic charge of matter field composites. In general one can
consider magnetic monopoles charged under $U(1)^N\subset U(N)$
with $H=(Q_1, ..., Q_N)$. Without loss of generality one can take
$Q_1\ge Q_2 \ge ... \ge Q_N$. Since elementary fields have unit
charges and transform in the fundamental of $SU(N)$, these
monopole operators correspond to Young Tableaux with $kQ_i$ boxes
in the $i^{th}$ row. For $k=1,2$ dressing composite vector
currents in the ${\bf 6}_{\pm 2}$ and scalar operators in the
${\bf 10}_{\pm 2}$ and ${\bf 10}^*_{\mp 2}$ (with
$\Delta_{\pm}=1,2$) with charge 2 monopole operators is crucial to
the enhancement of supersymmetry to $\cN = 8$ with full $SO(8)$
R-symmetry \cite{Klebanov:2009sg}. Monopole and anti-monopole
operators however appear in the spectrum even when $k\ge 3$ and no
(super)symmetry enhancement takes place \cite{Klebanov:2008vq,
Benna:2009xd}.

Before concluding this preliminary look, let us note that out of
the two $U(1)$ in the boundary CS theory only the Baryonic
$U(1)_B=U(1)_-$ is visible as a global symmetry, whose $\mZ_k$
subgroup is gauged, in the bulk description. The fate of the other
$U(1)$ is a sort of Higgs mechanism, under which $A_M \rightarrow
A_\mu$ and $C_{MNP} \rightarrow C_\mu \cJ_{ab}$ mix. Only the
combination $kA_\mu + N C_\mu$ remains massless and couples to
$U(1)_B$ while the orthogonal combination $NA_\mu - k C_\mu$
becomes massive by `eating' the (pseudo)scalar $\beta$ from $B_2 =
\beta \cJ$. A 5-brane instanton is thus expected to mediate
processes in which $k$ D0-branes transform into $N$ D4-branes
wrapped around $CP^2\subset \mC\mP^3$ \cite{Aharony:2008gk}.

\section{Compactification on $S^7$ revisited}

For the later use let us briefly review the mass spectrum of the
Freund-Rubin solution of $d=11$ supergravity on $S^7$
\cite{Sezgin:1983ik, Castellani:1984vv, VanNieuwenhuizen:1985be}.
The gravitino field  as well as all the fermions are set to zero,
the $AdS_4$ Riemann tensor and the three-form field strength are
given by: \bea
R_{\mu\nu\rho\sigma}=-4(g_{\mu\rho}(x)g_{\nu\sigma}(x)-g_{\mu\sigma}(x)g_{\nu\rho}(x))\\
F_{\mu\nu\rho\sigma}=3\sqrt{2}\sqrt{-\det g_{\mu\nu}(x)}\varepsilon_{\mu\nu\rho\sigma}
\eea
where $\varepsilon_{0123}=-1$.
The metric and the three form field with mixed indices vanish:
\bea
g_{\mu\alpha}=F_{\mu\nu\rho\alpha}=F_{\mu\nu\alpha\beta}=F_{\mu\alpha\beta\gamma}=0
\eea
and also
\bea
F_{\alpha\beta\gamma\delta}(y)=0\\
R_{\alpha \beta}=-6g_{\alpha \beta}(y)
\eea
$\mu, \nu, \rho =0,...,3$ are
$d=4$ indices, $\alpha, \beta, \gamma =1,...,7$ are internal indices.

Let us then consider fluctuations around the Freund-Rubin
solution. The linearized field equations are obtained by replacing
the background fields in the $d=11$ field equations by background
fields plus arbitrary fluctuations. An elegant and quite general
method to determine the complete mass spectrum on any coset
manifold relies on generalized harmonic expansion. In our case,
one expands the fluctuations in a complete set of spherical
harmonics of $S^7 = SO(8)/SO(7)$. The coefficient functions of the
spherical harmonics correspond to the physical fields in $d=4$. In
order to diagonalize the linearized equations it turns out to be
convenient to parameterize the fluctuations as follows: \bea
&& g_{\mu \nu}(x,y)=g_{\mu \nu}(x)+h_{\mu \nu}(x,y)\\
&& h_{\mu \nu}(x,y)=h_{\mu \nu}^\prime(x,y)-\frac{1}{2}g_{\mu \nu}(x)h_\alpha ^{\,\,\, \alpha}(x,y)\label{Weylresc}\\
&&g_{\alpha \beta}(x,y)=g_{\alpha \beta}(x)+h_{\alpha \beta}(x,y)\\
&& g_{\mu \alpha}(x,y)=h_{\mu \alpha}(x,y)\\
&& A_{\mu \nu \rho}(x,y)=A_{\mu \nu \rho}(x)+a_{\mu \nu \rho}(x,y)
\eea In particular the Weyl rescaled spacetime metric appears in
(\ref{Weylresc}) so as to put the $d=4$ Einstein action  in
canonical form. The spherical harmonic expansions of the
fluctuations of the metric and of the antisymmetric tensor fields
are given by: \bea
&& h_{(\mu \nu)}^\prime(x,y)=\sum H_{\mu \nu}^{N_1}(x)Y^{N_1}(y)\nn\\
&& h_{\mu \alpha}(x,y)=\sum B_\mu ^{N_7}(x)Y_\alpha ^{N_7}(y)+B_\mu ^{N_1}(x)D_\alpha Y^{N_1}(y)\nn\\
&& h_{(\alpha \beta)}(x,y)=\sum \phi ^{N_{27}}(x)Y_{(\alpha \beta)}^{N_{27}}(y)+\phi ^{N_7}
(x)D_{(\alpha}Y_{\beta)}^{N_7}(y)
+\phi ^{N_1}(x)D_{(\alpha}D_{\beta)}Y^{N_{1}}(y)\nn\\
&& h_\alpha ^{\,\,\,\alpha}(x,y)=\sum \pi^{N_1}(x)Y^{N_1}(y)\nn\\
&& A_{\mu \nu \rho}(x,y)=\sum a_{\mu\nu\rho}^{N_1}(x)Y^{N_1}(y)\nn\\
&& A_{\mu \nu \alpha}(x,y)=\sum a_{\mu \nu}^{N_7}(x)Y_\alpha ^{N_7}(y)+a_{\mu \nu}^{N_1}(x)D_\alpha Y^{N_1}(y)\nn\\
&& A_{\mu \alpha \beta}(x,y)=\sum a_\mu ^{N_{21}}(x)Y_{\alpha \beta}^{N_{21}}(y)+a_\mu ^{N_7}(x)D_{[\alpha}Y_ {\beta
]}^{N_7}\nn\\
&& A_{\alpha \beta \gamma}(x,y)=\sum a^{N_{35}}(x)Y_{\alpha \beta
\gamma}^{N_{35}}(y)+a^{N_{21}}(x)D_{[\alpha}Y_{\beta
\gamma]}^{N_{21}}(y) \eea All  superscripts $N_{{\bf r}}$ (${\bf
r}=1,7,21,27,35$) have infinite range, since they should provide a
basis for arbitrary fields on the 7-sphere. The index ${\bf r}$
specifies the $SO(7)$ representation of the corresponding
spherical harmonic. For example, $Y_{\alpha\beta\gamma}^{N_{35}}$
is in the third rank totally antisymmetric representation of
$SO(7)$ with dimension 35, while $Y_{(\alpha\beta)}^{N_{27}}$
is in the symmetric traceless 27-dimensional representation.
Derivatives of $Y$'s appear in the expansions since any tensor can
be decomposed into its transverse and longitudinal parts. After
fixing all local symmetries which do not correspond to gauge
invariances of the final $d=4$ theory and by choosing de Donder
type, $D^\alpha h_{(\alpha \beta)}(x,y)=0$, and Lorentz type,
$D^\alpha h_{\alpha \mu}(x,y)=0$, conditions the last term in
$h_{\mu \alpha}$ and the last two terms in $h_{(\alpha \beta)}$
drop out. To fix the local symmetries of the antisymmetric tensor
fields we choose the Lorentz conditions $D^\alpha
A_{\alpha\beta\gamma}(x,y)=D^\alpha
A_{\alpha\beta\mu}(x,y)=D^\alpha A_{\alpha\mu\nu}(x,y)=0$. As a
consequence, also these fields have only transverse harmonics
$a_{\mu\nu}^{N_1} (x)=a_\mu^{N_7}(x)=a^{N_{21}}(x)=0$.
Substituting the resulting expansions into the $d=11$ field equations,
the coefficients of each independent spherical harmonic yield the
$d=4$ field equations.

In the Einstein equation for $R_{\mu\nu}$ only $Y^{N_1}$ spherical
harmonics appear without derivatives. Thus there is only one field
equation, \ie one KK tower, for traceless symmetric tensors in
$AdS_4$.

Examining the Einstein equation for $R_{\alpha\beta}$ one can see
that the vector fields $B_\mu ^{N_7}$ are massive and transversal,
except for the lowest lying state corresponding to the Killing
vectors on $S^7$. The spin-0 fields $\phi ^{N_{27}}$ have a mass
matrix $\Delta _y+12$ ($\Delta _y$ is the Hodge-de Rham operator).
By a judicious gauge choice one can
eliminate $H_{\mu}^{N_1 \, \mu}$ in favour of $\pi ^{N_1}$ namely
$H_{\mu}^{N_1 \, \mu}=\frac{9}{7}\pi ^{N_1}$.

Collecting the coefficients of the spherical harmonics $Y_\alpha
^{N_7}$ and $D_\alpha Y^{N_1}$ in the Einstein equation for
$R_{\mu\alpha}$, one finds that the spin-1 spectrum consists of
linear combinations of $B_\mu ^{N_7}$ and $C_\mu ^{N_7}$ (from
$a_{\rho\sigma}^{N_7}$) and that one can eliminate the divergence
$D^\mu H^{N_1}_{\mu\nu}$ in favour of $\pi ^{N_1}$,
$a_{\rho\sigma\tau}^{N_1}$ except when $Y^{N_1}$ is a constant.

Similarly, inspecting the equations for $p$-form field strengths
($p=1,2,3,4$), one concludes that field expansions in
spherical harmonics can be chosen such that only the first terms
in the expansions survive with $Y$s being transversal and
traceless.

In particular, from the three-form field strength equation one
finds that $a_{\mu\nu\rho}^{N_1}=\varepsilon
_{\mu\nu\rho\lambda}D^\lambda \sigma^{N_1}$. This implies that the
divergence of $H_{\mu\nu}^{N_1}$ is proportional to a gradient.

From the four-form field strength equation one gets an equation
for $\Box _x \sigma^{N_1}$. Taking the trace of the equations for
$R_{\mu\nu}$ and $R_{\alpha\beta}$, an equation involving $\Box _x
\sigma^{N_1}$ and $\Box _x H_\mu ^{N_1\mu}$ arises. Resolving the
mixing between $a_{\mu\nu\rho}^{N_1}$ and $H_\mu ^{N_1\mu}$
produces to independent combinations and as many KK towers of
scalars.

From the two-form field strength equation one finds $D^\mu
a_{\mu\nu}^{N_7}=0$, which implies
$a_{\mu\nu}^{N_7}=\varepsilon_{\mu\nu}^{\,\,\,\,\,\,\,\rho\sigma}D_\rho
C_\sigma ^{N_7}$. Using one of the three-form field strength
equations one finds that $C^{N_7}_\mu$ and $B^{N_7}_\mu$ mix.
Resolving the mixing one finds two KK towers, one of which starts
with a massless vector corresponding to the internal Killing
vectors of $S^7$.

After diagonalizing the bosonic field equations one obtains the
mass spectrum summarized in Table \ref{bosons}.
\begin{table}
\begin{center}
  \begin{tabular}{ | l | l | l | l | l | l | l |}\hline
Spin & Field & $SO(7)$ & $SO(8)$ &  $4(ML)^2$ & $\Delta$ & $\ell$
\\ \hline
$2^+$ & $h_{(\mu \nu)}^\prime$ & $N_1$ & $(\ell,0,0,0)$ & $\ell (\ell +6)$ & $\Delta=\frac{\ell}{2}+3$  & $ \ell\geq 0$ \\
\hline
$1_1^-$ & $h_{\mu \alpha}$ & $N_7$ & $(\ell,1,0,0)$ & $\ell(\ell +2)$ & $\Delta=\frac{\ell}{2}+2$ & $\ell\geq 0$ \\
$1_2^-$ & $A_{\mu \nu \alpha}$ & $N_7$ & $(\ell-2,1,0,0)$ & $(\ell +6)(\ell +4)$ & $\Delta=\frac{\ell}{2}+4$ & $\ell\geq
2$\\
\hline
$1^+$ & $A_{\mu \alpha \beta}$ & $N_{21}$ &
$(\ell-1,0,1,1)$ & $(\ell +2)(\ell +4)$ &
$\Delta=\frac{\ell}{2}+3$ & $\ell\geq 1$\\ \hline
$0_1^+$ & $A_{\mu \nu \rho}$ & $N_1$ & $(\ell+2,0,0,0)^*$ & $(\ell +2)(\ell
-4)$ & $\Delta=\frac{\ell}{2}+1$ & $\ell\geq 0$\\
$0_2^+$ & $h_{\alpha \alpha},\,h_{\lambda \lambda}^\prime$ & $N_1$ & $(\ell-2,0,0,0)$ & $(\ell +10)(\ell +4)$ &
$\Delta=\frac{\ell}{2}+5$ & $\ell\geq 2$\\ \hline
$0_3^+$ & $h_{(\alpha \beta)}$ & $N_{27}$ & $(\ell-2,2,0,0)$ & $\ell (\ell
+6)$ & $\Delta=\frac{\ell}{2}+3$ & $\ell\geq 2$\\ \hline
$0_1^-$ &$A_{\alpha \beta \gamma}$ & $N_{35}$ & $(\ell,0,2,0)$ & $(\ell
-2)(\ell +4)$ & $\Delta=\frac{\ell}{2}+2$ & $\ell\geq 0$\\ \hline
$0_2^-$ & $A_{\alpha \beta \gamma}$ & $N_{35}$ & $(\ell-2,0,0,2)$
& $(\ell +8)(\ell +2)$ & $\Delta=\frac{\ell}{2}+4$ & $\ell\geq
2$\\ \hline
    \end{tabular}
\caption{Bosonic KK towers after compactification on $S^7$}
\end{center}
\label{bosons}
\end{table}
The resulting bosonic spectrum includes the massless graviton,
${\bf 28}$ massless vectors of $SO(8)$, corresponding to a
combination of $B_\mu$ (in $h_{\mu\a}$) and $C_\mu$ (in
$A_{\mu\nu\a}$), ${\bf 35}_v$ scalars ($\Delta =1$) and ${\bf
35}_s$ ($\Delta =2$) pseudoscalars with $(ML_{AdS})^2 = -2$. In
the supergravity literature \cite{Sezgin:1983ik,
Castellani:1984vv, VanNieuwenhuizen:1985be} masses of scalars are
often shifted by $-R/6$ so that $(ML_{AdS})^2 \rightarrow
(\tilde{M}L_{AdS})^2 = (ML_{AdS})^2 + 2$. The 70 (pseudo)scalars
in the $\cN = 8$ supergravity multiplet are `massless' in the
sense that $(\tilde{M}L_{AdS})^2=0$. Moreover, there are three
families of scalars and two families of pseudoscalar excitations.
Three of them ($0_2^+$, $0_3^+$ and $0_2^-$) contain only states
with positive mass square and correspond to irrelevant operators
in the dual CFT. The remaining families $0_1^+$ and $0_1^-$
contain states with positive, zero and negative mass squared
corresponding to irrelevant, marginal and relevant operators,
respectively.

A similar analysis can be performed for fermionic fluctuations. In
Table \ref{fermions} we summarize the fermionic mass spectrum.
\begin{table}
\begin{center}
 \begin{tabular}{ | l | l | l | l | l |}
    \hline
Spin &  $SO(8)$ &  $ 4(ML)^2$ & $\Delta$ & $\ell$ \\ \hline
$(\frac{3}{2})_1$ & $(\ell,0,0,1)$ & $ (\ell +2)^2$ & $\Delta=\frac{\ell}{2}+\frac{5}{2}$  & $ \ell\geq 0$ \\
\hline
$(\frac{3}{2})_2$ & $(\ell -1,0,1,0)$ & $ (\ell +4)^2$ & $\Delta=\frac{\ell}{2}+\frac{7}{2}$  & $ \ell\geq 1$ \\
\hline
$(\frac{1}{2})_1$ & $(\ell +1,0,1,0)^*$ & $ \ell ^2$ & $\Delta=\frac{\ell}{2}+\frac{3}{2}$  & $ \ell\geq 0$ \\
\hline
$(\frac{1}{2})_2$ & $(\ell -1,1,1,0)$ & $ (\ell +2) ^2$ & $\Delta=\frac{\ell}{2}+\frac{5}{2}$  & $ \ell\geq 1$ \\
\hline
$(\frac{1}{2})_3$ & $(\ell -2,1,0,1)$ & $ (\ell +4)^2$ & $\Delta=\frac{\ell}{2}+\frac{7}{2}$  & $ \ell\geq 2$ \\
\hline
$(\frac{1}{2})_4$ & $(\ell -2,0,0,1)$ & $ (\ell +6) ^2$ & $\Delta=\frac{\ell}{2}+\frac{9}{2}$  & $ \ell\geq 2$ \\
\hline
 \end{tabular}
 \caption{Fermionic KK towers after compactification on $S^7$}
\end{center}
\label{fermions}
\end{table}

The KK spectrum does not include the states with $*$ for $\ell
=-1$, since they do not propagate in the bulk but live on the
conformal boundary of $AdS_4$. They correspond to the singleton
representation of $Osp(8|4)$ that consists of $8_v$ bosons $X^i$
with $\Delta =\frac{1}{2}$, $(ML)^2=-\frac{5}{4}$ and $8_c$
fermions $\psi ^{\dot{a}}$ with $\Delta =1$, $ML=\frac{1}{2}$,
both at the unitary bound.

The KK excitations on $S^7$ can be put in one-to-one
correspondence with `gauge-invariant' composite operators on the
boundary. The dictionary for bosonic operators schematically
reads: \bea && s=2^+ \quad  \quad T^{i_1...i_\ell}_{\mu \nu,
\Delta = {\ell \over 2} + 3}= (\partial _\mu X_i
\partial _\nu X^i+\bar{\psi}\gamma _\mu \partial _\nu
\psi)X^{i_1}...X^{i_\ell}\\
&&  s=1^-_1 \quad \quad J_{\mu, \Delta = {\ell \over 2} + 2
}^{[ij]i_1...i_\ell}= (X^{[i}{\partial} _\mu X^{j]}+ \bar{\psi}
\Gamma^{ij}\gamma_\mu
\psi)X^{i_1}...X^{i_\ell}\\
&& s=1^-_2 \quad  \quad J^{[ij]i_1...i_{\ell-2}}_{\mu, \Delta =
{\ell \over 2} + 4} =\partial _\mu X_i\partial _\nu X^i \bar{\psi}
\gamma ^\nu \Gamma ^{ij}\psi X^{i_1}...X^{i_{\ell-2}}\\
&& s=1^+ \quad  \quad J^{a\dot{b}i_1...i_{\ell-1}}_{\mu,\Delta =
{\ell \over 2} + 3}= \bar{\psi}\Gamma _{jk}\partial
_\mu \psi (X_i \Gamma ^{ijk})^{a\dot{b}}X^{i_1}...X^{i_{\ell-1}} \\
&& s=0_1^+ \quad  \quad \Phi ^{ij i_1...i_\ell}_{\Delta = {\ell
\over 2} + 1}= X^{i} X^j
X^{i_1}...X^{i_\ell}\\
&& s=0_2^+ \quad  \quad \Phi ^{i_1...i_{\ell-2}} _{\Delta = {\ell
\over 2} + 5}=
\partial _\mu X^i \partial _\nu X_i
\bar\psi\gamma ^\mu \partial ^\nu \psi X^{i_1}...X^{i_{\ell-2}} \\
&& s=0_3^+ \quad  \quad \Phi ^{[ij][kl]i_1...i_{\ell-2}}_{\Delta =
{\ell \over 2} + 3}= (\bar{\psi} \Gamma ^{ij}
\gamma _\mu \psi X^{[k} \partial ^\mu X^{l]})X^{i_1}...X^{i_{\ell-2}}\\
&& s=0_1^- \quad  \quad
\Phi^{(\dot{a}\dot{b})i_1...i_\ell}_{\Delta = {\ell \over 2} + 2}=
\bar\psi ^{\dot{a}}
\psi ^{\dot{b}} X^{i_1}...X^{i_\ell}\\
&& s=0_2^- \quad \quad \Phi^{(a b)i_1...i_{\ell-2}}_{\Delta =
{\ell \over 2} + 4 }= (\Gamma ^{ijkl})^{ab} X_i\partial ^\mu
X_j\bar{\psi}\Gamma _{kl}\partial _\mu \psi
X^{i_1}...X^{i_{\ell-2}} \eea A similar dictionary can be compiled
for fermions.

\section{Polynomial representations for $SO(8)$ and $U(4)$}
\label{PolyRep}
In order to decompose KK harmonics on $S^7=SO(8)/SO(7)$ into KK
harmonics on $\mC\mP^3=U(4)/U(3)\times U(1)$, we will present the
construction of arbitrary representations of $SO(8)$ in the space
of polynomials of $12$ variables. The latter are the coordinates
of the subgroup $Z_+^{SO(8)}$ generated by the raising operators
of $SO(8)$. We will then describe a technique which allows to
identify which of the above polynomials correspond to highest
weight states of representations of $U(4)\subset SO(8)$. The method we use
is quite standard in representation theory of Lie groups (see \eg Chapter 16
of \cite{Zhelobenko}).

It is convenient to start with $SO(8,\mC)$ defined as the group of
$8\times 8$ complex matrices which leave invariant the quadratic
form $X^TC^{(8)}X$, where $X$ is a complex (column) vector whose
components will be enumerated as $X^1,X^2,X^3,X^4,X^{\tilde
4},X^{\tilde 3},X^{\tilde 2},X^{\tilde 1}$ and $C^{(8)}$ is an
$8\times 8$ matrix with $1$'s on SW-NE (anti)diagonal: \be
C^{(8)}_{ij}=C^{(8)}_{{\tilde i}{\tilde j}}=0, \hspace{1cm}
C^{(8)}_{i{\tilde j}}=C^{(8)}_{{\tilde j}i} = \delta_{ij},
\hspace{0.5cm} i,j=1,2,3,4 \ee By definition all matrices  $g\in
SO(8)$ satisfy the condition $g^TC^{(8)}g=C^{(8)}$. Eventually, in
order to select the compact real form $SO(8)$ of our interest, one
should identify the coordinates $X^{{\tilde i}}$ with ${\bar X}^i$
(bar means complex conjugate). A generic $SO(8)$ matrix $g$  can
be (uniquely) decomposed as (Gauss decomposition): \be g=\zeta
\lambda z, \ee where $\zeta \in Z_ -$, $z \in Z_+ $, $\lambda \in
\Lambda $ with $Z_+$ ($Z_-$) being the subgroup of lower (upper)
triangular matrices with $1$'s on the diagonal and $\Lambda$ is
the subgroup of diagonal matrices (Cartan subgroup). Let's set
$\lambda =Diag(\lambda_1,
\lambda_2,\lambda_3,\lambda_4,\lambda_4^{-1},\lambda_3^{-1},
\lambda_2^{-1},\lambda_1^{-1})$. We will realize the irreducible
representations of the group $SO(8)$ on some spaces of functions
defined on it. In particular, the role of the highest weight
vector will be played by the function : \be \alpha
(g)=\lambda_1^{m_1} \lambda_2^{m_2} \lambda_3^{m_3}
\lambda_4^{m_4} \label{charm} \ee where $m_1\ge m_2\ge m_3\ge
|m_4| $ ($m_i$ are either all integers or all half-integers)
uniquely characterize the irrep. The eigenvalues $\lambda_i$ can
be expressed  in terms of the matrix elements of $g$ explicitly:
 \be
 \lambda_p=\frac{\Delta_p}{\Delta_{p-1}},\hspace{0.5cm} p=1,2,3,4
 \ee
where $\Delta_0=1$ and $\Delta_p$, $p=1,2,3,4$ are the diagonal
minors \bea
 \Delta_p=\left|
\begin{array}{ccc}
g_{11}&\cdots &g_{1p}\\
\vdots&\cdots &\vdots\\
g_{p1}&\cdots &g_{pp}\\
\end{array}
 \right|.
 \eea
Introducing the notation $S_-=\frac{\Delta_3}{\sqrt{\Delta_4}}$,
$S_+=\sqrt{\Delta_4}$ (it is easy to see that $S_{+,-}$
polynomially depend on the matrix elements of $g$) we can rewrite
eq. (\ref{charm}) as \be \alpha (g)=\Delta_1^{\ell_1} \Delta_2^{\ell_2}
S_-^{\ell_3} S_+^{\ell_4} \label{charm1} \ee where $\ell_1=m_1-m_2$,
$\ell_2=m_2-m_3$, $\ell_3=m_3-m_4$ and $\ell_4=m_3+m_4$ are non-negative
integers commonly referred as the Dynkin labels of the irrep.
Consider the space ${\cal R}_\alpha$ of all linear combinations of
the functions $\alpha(gg_0)$,  $g_0\in SO(8)$. $SO(8)$ is
represented in ${\cal R}_\alpha$ simply by the right multiplication of
the argument. As already mentioned the function $\alpha(g)$ plays the role of the
highest weight state. For any function $f(g)\in {\cal R}_\alpha $
we have $f(\zeta\lambda z)=\alpha (\lambda) f(z)$ which shows that
to restore its full $g$-dependence it is sufficient to only know
the values the function assumes on the subgroup $Z_+$. This is why
actually we get representation on a space of functions of $z$, in fact
polynomials due to the polynomial dependence on $g$ of $\alpha(g)$
mentioned earlier.

There is an elegant way to characterize this space of polynomials.
Consider the four raising generators corresponding to the simple
roots\bea
e_1=E_{12}-E_{{\tilde 2}{\tilde 1}}; \hspace{0.5cm}
e_2=E_{23}-E_{{\tilde 3}{\tilde 2}}\nonumber \\
e_-=E_{34}-E_{{\tilde 3}{\tilde 4}}; \hspace{0.5cm}
e_+=E_{3{\tilde 4}}-E_{{\tilde 4}3} \label{risinggen} \eea where
$E_{pq}$ denotes  the $8\times 8$ matrix whose only non-zero entry
$1$ is at the position $(p,q)$. Denote their  {\it left} action on
${\cal R}_\alpha $ by ${\cal D}_1$, ${\cal D}_2$, $ {\cal D}_-$,
${\cal D}_+$. It is not difficult to prove that \bea
{\cal D}_1^{\ell_1 +1}\alpha(g)=0\nonumber\\
{\cal D}_2^{\ell_2 +1}\alpha(g)=0\nonumber\\
{\cal D}_-^{\ell_3 +1}\alpha(g)=0\nonumber\\
{\cal D}_+^{\ell_4 +1}\alpha(g)=0. \eea The key observation is
that the same equations are valid also for arbitrary functions
$f\in {\cal R}_\alpha $, since they are all generated by
$\alpha(g)$ through right multiplications which commute with left
multiplications. Below we will use a convenient explicit
parametrization of $Z_+\subset SO(8)$ in terms of two $4\times 4$
matrices $\eta $ and $a$ \bea \eta= \left(
\begin{array}{cccc}
1&\eta_{12} &\eta_{13}&\eta_{14}\\
0&1 &\eta_{23}&\eta_{24}\\
0&0 &1&\eta_{34}\\
0&0 &0&1\\
\end{array}
 \right);\hspace{0.5cm}
a= \left(
\begin{array}{cccc}
a_{14}&a_{13} &a_{12}&0\\
a_{24}&a_{23}&0&-a_{12}\\
a_{34}&0&-a_{23}&-a_{13}\\
0&-a_{34}&-a_{24}&-a_{14}
\end{array}
 \right).
\eea Let us further introduce the $8\times 8$ matrices which in
$2\times 2$ block form read \bea
 z_0=
\left(
\begin{array}{cc}
\eta &0\\
0&{\tilde \eta}\\
\end{array}
\right)
 ;\hspace{0.5cm}
z^\prime= \left(
\begin{array}{cc}
1&a\\
0&1
\end{array}
 \right),
\eea
where
\bea
{\tilde \eta}=
\left(
\begin{array}{cccc}
1& -\eta _{34} & -\eta _{24}+\eta _{23} \eta _{34} & -\eta _{14}+\eta _{12} \eta _{24}+\eta _{13} \eta _{34}-\eta _{12}
\eta _{23} \eta _{34} \\
0& 1 & -\eta _{23} & -\eta _{13}+\eta _{12} \eta _{23} \\
0& 0 & 1 & -\eta _{12}\\
0&0&0&1
\end{array}
 \right).
 \label{etaa}
\eea An arbitrary $z\in Z_+$ can be (uniquely) represented as
\be z=z^\prime z_0. \label{etaadecomposition} \ee Left
multiplication by raising generators (\ref{risinggen}) induces
infinitesimal motion on the parameters $a$, $\eta$. A
straightforward algebra shows that \eg \bea (1+\epsilon
e_1)z(a,\eta)=z(a+\delta  a,\eta+\delta\eta)+O(\epsilon^2), \eea
where the non-trivial variations are \bea \delta\eta _{12}=
\epsilon ,\,\,\delta \eta _{13}=\epsilon \eta _{23},\,\,\delta\eta
_{14}=\epsilon \eta _{24},\,\, \delta a_{13}=\epsilon a_{23},\,\,
\delta a_{14}=\epsilon a_{24}.\nonumber \eea Similarly examining the
remaining three generators we find \bea {\cal
D}_1&=&\partial_{\eta_{12}}+\eta_{23}
\partial_{\eta_{13}}+a_{23}\partial_{a_{13}}+
a_{24}\partial_{a_{14}} \nonumber\\
{\cal D}_2&=&\partial_{\eta_{23}}+\eta_{34} \partial_{\eta_{24}}+a_{13}\partial_{a_{12}}+
a_{34}\partial_{a_{24}} \nonumber\\
{\cal D}_-&=&\partial_{\eta_{34}}+a_{14}\partial_{a_{13}}+
a_{24}\partial_{a_{23}} \nonumber\\
{\cal D}_+&=&\partial_{a_{34}}. \eea
Thus any irreducible
representation of $SO(8)$ is realized on the space of polynomials
 of $12$ variables $a$, $\eta$ subject to the constraints
\bea
&&\left(\partial_{\eta_{12}}+\eta_{23} \partial_{\eta_{13}}+a_{23}\partial_{a_{13}}+
a_{24}\partial_{a_{14}}\right)^{\ell_1 +1}f(a,\eta)=0 \nonumber\\
&&\left(\partial_{\eta_{23}}+\eta_{34} \partial_{\eta_{24}}+a_{13}\partial_{a_{12}}+
a_{34}\partial_{a_{24}}\right)^{\ell_2 +1}f(a,\eta)=0  \nonumber\\
&&\left(\partial_{\eta_{34}}+a_{14}\partial_{a_{13}}+
a_{24}\partial_{a_{23}}\right)^{\ell_3 +1}f(a,\eta)=0  \nonumber\\
&&\left(\partial_{a_{34}}\right)^{\ell_4 +1}f(a,\eta)=0 .
\label{indicatorsystem}
\eea

Note that the constant polynomial always satisfies
(\ref{indicatorsystem}) and corresponds to the highest weight
state. Considering right multiplication it is not difficult to
find explicit expressions for the generators of $SO(8)$ as
operators acting on the space of polynomials. For our later
proposes let us specify how the diagonal part $\Lambda\subset
SO(8)$ is represented. Since \be z(a,\eta)\lambda=\lambda
\lambda^{-1}z(a,\eta)\lambda =\lambda z(a^\prime ,\eta^\prime),
\ee where \bea a^\prime_{ij}=\lambda_j^{-1}\lambda_i^{-1}a_{ij};
\hspace{0.5cm} \eta^\prime_{ij}=\lambda_j\lambda_i^{-1}\eta_{ij}
\eea we simply get \be \lambda \circ
f(a,\eta)=\lambda_1^{m_1}\lambda_2^{m_2}\lambda_3^{m_3}\lambda_4^{m_4}f(a^\prime
,\eta^\prime) \ee Notice that the variable $a_{ij}$ shifts the
weights as $m_i\rightarrow m_i-1$, $m_j\rightarrow m_j-1$
while the variable $\eta_{ij}$ shifts them as $m_i\rightarrow
m_i-1$, $m_j\rightarrow m_j+1$.

Consider now the $GL(4,C)\subset SO(8,C)$ subgroup whose
off-diagonal blocks in $2\times 2$ block notation are zero. This
subgroup does not mix the coordinates $X^i$ with $X^{{\tilde i}}$
and after restriction to the real sector it becomes the subgroup
$U(4)\subset SO(8)$.

In other words, for the reduction from $S^7$ to $S^7/\mZ_k$ or
$\mC\mP^3 \ltimes S^1$ we are interested in, the decomposition
$SO(8)\rightarrow SO(6)\times SO(2)$ is given by the embedding \be
{\bf 8_v}(1,0,0,0) \rightarrow {\bf 4}_{+1}[0,1,0] + {\bf
4^*}_{-1}[0,0,1] \label{8v} \ee where $(\ell _1,\ell _2,\ell
_3,\ell _4)$ and $[k,l,m]$ denote $SO(8)$ and $SO(6)$ Dynkin
labels respectively. As a result, for the Adjoint representation
one has \be {\bf 28}(0,1,0,0) \rightarrow {\bf 15}_{0}[0,1,1] +
{\bf 1}_{0}[0,0,0] + {\bf 6}_{+2}[1,0,0] + {\bf 6}_{-2}[1,0,0] \ee
while \be {\bf 8_s}(0,0,0,1) \rightarrow {\bf 6}_0[1,0,0] + {\bf
1}_{+2}[0,0,0] + {\bf 1}_{-2}[0,0,0] \label{8s} \ee \be {\bf
8_c}(0,0,1,0)\rightarrow {\bf 4}_{-1}[0,1,0] + {\bf
4^*}_{+1}[0,0,1] \label{8c} \ee for the spinorial representations.

Our goal is to identify the highest weight states of this subgroup
inside the space of polynomials of a given representation of
$SO(8)$. It is evident from the decomposition
(\ref{etaadecomposition},\ref{etaa})  that the right action by the
raising operators of $GL(4)$ subgroup $e_1$, $e_2$, $e_-$ (see eq.
(\ref{risinggen})) shifts the parameters $\eta$ and leave the
parameters $a$ untouched. Thus, in order to be a highest weight
state, a polynomial, besides satisfying the equations
(\ref{indicatorsystem}) should be independent of $\eta$. The {\it
indicator} system for the highest weight states becomes \bea
&&\left(a_{23}\partial_{a_{13}}+
a_{24}\partial_{a_{14}}\right)^{\ell_1 +1}f(a)=0 \nonumber\\
&&\left(a_{13}\partial_{a_{12}}+
a_{34}\partial_{a_{24}}\right)^{\ell_2 +1}f(a)=0  \nonumber\\
&&\left(a_{14}\partial_{a_{13}}+
a_{24}\partial_{a_{23}}\right)^{\ell_3 +1}f(a)=0  \nonumber\\
&&\left(\partial_{a_{34}}\right)^{\ell_4 +1}f(a)=0 .
\label{hwindicatorsystem}
\eea
Solving these equations one can fully decompose KK
harmonics on $S^7$ into KK harmonics of $\mC\mP^3\times S^1$ which is our
next task.

\section{From $S^7$ to $\mC\mP^3\ltimes S^1$}

$S^7$ is a $U(1)$ bundle over $\mC\mP^3$. The $\mC\mP^3$ solution of the
$d=10$ theory can be obtained from the $S^7$ solution of the
$d=11$ theory by Hopf fibration, \ie keeping only $U(1)$ invariant
states \cite{Nilsson:1984bj}. The compactification on $\mC\mP^3$ of the
$d=10$ theory yields a four dimensional theory with $\cN=6$
supersymmetry and with gauge group $SO(6)\times SO(2)$.

The truncation from $S^7$ to $\mC\mP^3 \ltimes S^1$ cannot be thought
of as spontaneous (super)symmetry breaking and one has to really
discard the states that are projected out by $\mZ_k$ or $SO(2)$ for
$k\rightarrow \infty$ even if it acts freely. In particular we
will later check that no Higgsing can account for the breaking of
$SO(8)$ to $SO(6)\times SO(2)$ but rather the coset vectors are
dressed with monopole operators and become massive for $k\neq 1,2$
\cite{Aharony:2008gk, Aharony:2008ug, Benna:2008zy, Benna:2009xd,
Klebanov:2009sg}.

Let us start with the KK towers of bosons. Using the procedure
described in the previous section or otherwise, for scalar
spherical harmonics with Dynkin labels $(\ell,0,0,0)$ one finds as
independent polynomials $\{a_{14}^m \,| \,m=0,...,\ell\}$. Thus
the following decomposition holds: \bea N_1: \quad (\ell,0,0,0)\rightarrow
\oplus [0,\ell -m,m]_{\ell -2m} \eea where the subscript is the
$SO(2)$ charge $Q$ of the appropriate representation.

For vector spherical harmonics with $SO(8)$ Dynkin labels
$(\ell-2,1,0,0)$ one gets $\{a_{12}a_{14}^m , \, a_{24}a_{14}^m
,\, (a_{13}a_{24}-a_{14}a_{23})a_{14}^m , \, a_{14}^m \, | \,
m=0,...,\ell\}$ as independent polynomials. The $SO(8)$
representation decomposes into $SO(6)$ representations as: \bea N_7: \quad
(\ell,1,0,0)\rightarrow && \oplus[0,\ell -m,m]_{\ell -2m}
\oplus [0,\ell -m+1,m+1]_{\ell -2m }\nonumber\\
&& \oplus [1,\ell -m,m]_{\ell -2m-2} \oplus [1,\ell -m,m]_{\ell
-2m+2} \eea One obtains the decomposition of the representation
$(\ell -2, 1, 0, 0)$ from the previous one by shifting $\ell$ to
$\ell-2$. In what follows we will simply omit the decompositions
which differ by shifts of the parameter $\ell$.

For two-form spherical harmonics with $SO(8)$ Dynkin labels $(\ell
-1,0,1,1)$ one finds $\{a_{14}^m , \, a_{23}a_{14}^m , \,
a_{34}a_{14}^m , \, a_{23}a_{34}a_{14}^m ,\, (a_{34}a_{12} -
a_{13}a_{24})a_{14}^m ,\, a_{23}(a_{23}a_{14} + a_{34}a_{12} -
a_{13}a_{24})a_{14}^m),\, a_{13}a_{14}^n , \, a_{34}a_{13}a_{14}^n
, \, (a_{34}a_{12}-a_{13}a_{24})a_{13}a_{14}^n \, | \,
m=0,...,\ell -1,\, n=0,..., \ell -2\}$ as independent polynomials.
One then finds the following decomposition: \bea N_{21}: \quad (\ell
-1,0,1,1)\rightarrow &&\oplus [0,\ell -m,m]_{\ell -2m-4}
\oplus [0,\ell -m-1,m+1]_{\ell -2m+2}\nonumber\\
&&\oplus [1,\ell -m,m]_{\ell -2m-2}
\oplus [1,\ell -m-1,m+1]_{\ell -2m}\nonumber\\
&&\oplus [0,\ell -m,m]_{\ell -2m}
\oplus [0,\ell -m-1,m+1]_{\ell -2m-2}\nonumber\\
&&\oplus [1,\ell -n-2,n]_{\ell -2n-4}
\oplus [2,\ell -n-2,n]_{\ell -2n-2}\nonumber\\
&&\oplus [1,\ell -n-2,n]_{\ell -2n}
\eea
The decomposition of the KK towers corresponding to $0_1^+$ and $0_2^+$
can be found from the decomposition of $2^+$ via
appropriate shifts.

For second rank symmetric traceless harmonics with Dynkin labels
$(\ell -2,2,0,0)$ the polynomials are: $\{a_{14}^m $, $
a_{12}a_{14}^m $, $a_{12}(a_{23}a_{14}-a_{13}a_{24})a_{14}^m$, $
a_{12}^2a_{14}^m$, $a_{12}a_{24} a_{14}^m $,$a_{24}a_{14}^m$,
$a_{24}(a_{23}a_{14}-a_{13}a_{24})a_{14}^m $, $(a_{13}a_{24}-
a_{14}a_{23})a_{14}^m$, $(a_{14}a_{23}-a_{13}a_{24})^2a_{14}^m$,
$a_{24}^2 a_{14}^m $, $|\, m=0,...,\ell -2\}$. The $SO(6)$
representations decomposed from $SO(8)$'s are: \bea N_{27}: \quad
(\ell -2,2,0,0)\rightarrow && \oplus [2,\ell -m-2,m]_{\ell -2m+2}
\oplus [1,\ell -m-2,m]_{\ell -2m}\nonumber\\
&& \oplus [1,\ell -m-2,m]_{\ell -2m-4}
\oplus [0,\ell -m-1,m+1]_{\ell -2m-2}\nonumber\\
&& \oplus [0,\ell -m-2,m]_{\ell -2m-2}
\oplus [1,\ell -m-1,m+1]_{\ell -2m}\nonumber\\
&& \oplus [1,\ell -m-1,m+1]_{\ell -2m-4}
\oplus [2,\ell -m-2,m]_{\ell -2m-2}\nonumber\\
&& \oplus [2,\ell -m-2,m]_{\ell -2m-6} \oplus [0,\ell
-m,m+2]_{q=\ell -2m-2}\nn \\ \eea

For the three-form spherical harmonic with $SO(8)$ Dynkin labels $(\ell ,0,2,0)$ one finds
$\{(a_{14}^m + a_{23} a_{14}^m + a_{23}^2 a_{14}^m) ,\,
a_{13}(a_{14}^n + a_{23} a_{14}^n) ,\, a_{13}^2 a_{14}^p\, | \, m=0,...,\ell ,\, n=0,...,\ell -1 , \, p=0,...,\ell -2\}$
polynomials. The representation $(\ell ,0,2,0)$ decomposes as:
\bea
N_{35}: \quad (\ell ,0,2,0)\rightarrow && \oplus [0,\ell -m,m+2]_{\ell -2m+2}
\oplus [0,\ell -m+1,m+1]_{\ell -2m}\nonumber\\
&& \oplus [0,\ell -m+2,m]_{\ell -2m-2}
\oplus [1,\ell -n-1,n+1]_{\ell -2n}\nonumber\\
&& \oplus [1,\ell -n,n]_{\ell -2n-2}
\oplus [2,\ell -p-2,p]_{\ell -2p-2}
\eea

For the three-form spherical harmonic with $SO(8)$ Dynkin labels
$(\ell -2,0,0,2)$ one has $\{(a_{14}^m ,\, (a_{14} a_{23} + a_{12}
a_{34} - a_{13} a_{24})a_{14}^m ,\, (a_{12} a_{34} - a_{13} a_{24}
+ a_{14} a_{23})^2 a_{14}^m ,\, a_{34} a_{14}^m ,\, $ $a_{34}
(a_{24} a_{13} - a_{34} a_{12} - a_{14} a_{23}) a_{14}^m ,\,
a_{34}^2 a_{14}^m \, | \, m=0,...,\ell -2 )$ and the following
decomposition: \bea N_{35}^\prime : \quad (\ell -2,0,0,2)\rightarrow && \oplus [0,\ell
-m-2,m]_{\ell -2m-2}
\oplus [0,\ell -m-2,m]_{\ell -2m+2}\nonumber\\
&& \oplus [1,\ell -m-2,m]_{\ell -2m-4}
\oplus [1,\ell -m-2,m]_{\ell -2m}\nonumber\\
&& \oplus [2,\ell -m-2,m]_{\ell -2m-2} \oplus [0,\ell
-m-2,m]_{\ell -2m-6}\nn\\ \eea

Let us now consider the fermionic KK towers. There are two
gravitini in the $SO(8)$ representations $(\ell, 0,0,1)$ and
$(\ell-1, 0,1,0)$.

For the $SO(8)$ representation $(\ell, 0,0,1)$ one finds
$\{a_{14}^m , \, (a_{14}a_{23}+a_{12}a_{34}-a_{13}a_{24})a_{14}^m
, \,a_{34}a_{14}^m \, | \, m=0,..,\ell\}$ as polynomials and the
following decomposition holds  \bea (\ell ,0,0,1)\rightarrow \oplus
[0,\ell -m,m]_{\ell -2m+2} \oplus [0,\ell -m,m]_{\ell -2m-2}
\oplus [1,\ell -m,m]_{\ell -2m} \eea

For the $SO(8)$ representation $(\ell-1, 0,1,0)$ the independent
polynomials are $\{ a_{14}^m ,\, a_{23}a_{14}^m ,\,
a_{13}a_{14}^{m'} \, | \, m=0,...,\ell -1, \, {n}=0,...,\ell
-2\}$ and is decomposed as: \bea
(\ell -1,0,1,0)\rightarrow &&\oplus [0,\ell -m-1,m+1]_{\ell -2m} \oplus [0,\ell -m,m]_{\ell -2m-2}\nonumber\\
&& \oplus [1,\ell -{n}-2,{n}]_{\ell -2{n}-2} \eea There are
other fermions in the representations $(\ell+1,0,1,0)$,
$(\ell-2,0,0,1)$, $(\ell-1,1,1,0)$ and $(\ell-2,1,0,1)$.

For the $SO(8)$ representation $(\ell-1,1,1,0)$ the polynomials
have the form $\{a_{14}^m ,\, a_{23}a_{14}^m ,\, a_{23}(a_{13}a_{24}-a_{14}a_{23})a_{14}^m ,\, a_{24}a_{14}^m ,\,
a_{13}a_{24}a_{14}^m ,\, a_{23}a_{24}a_{14}^m ,\, a_{12}a_{14}^m,\,$ $a_{12}a_{23}a_{14}^m, a_{13}a_{14}^n ,\,
a_{13}(a_{13}a_{24}-a_{23}a_{14})a_{14}^n ,\,
a_{12}a_{13}a_{14}^n\, |\, m=0,...,\ell -1, n=0,...,\ell-2\}$ and
one has the following decomposition: \bea (\ell
-1,1,1,0)\rightarrow && \oplus [1,\ell -m-1,m+1]_{\ell -2m+2}
\oplus [1,\ell -m,m]_{\ell -2m}\nonumber\\
&& \oplus [1,\ell -m,m]_{\ell -2m-4}
\oplus [0,\ell -m,m+2]_{\ell -2m}\nonumber\\
&& \oplus [1,\ell -m-1,m+1]_{\ell -2m-2}
\oplus [0,\ell -m+1,m+1]_{\ell -2m-2}\nonumber\\
&& \oplus [0,\ell -m-1,m+1]_{\ell -2m}
\oplus [0,\ell -m,m]_{\ell -2m-2}\nonumber\\
&& \oplus [2,\ell -n-2,n]_{\ell -2n}
\oplus [2,\ell -n-2,n]_{\ell -2n-4}\nonumber\\
&& \oplus [1,\ell -n-2,n]_{\ell -2n-2}
\eea

Finally for the $SO(8)$ representation $(\ell-2,1,0,1)$ the
polynomials have the form $\{ a_{14}^m ,\,
(a_{14}a_{23}-a_{13}a_{24})a_{14}^m ,\,
(a_{13}a_{24}-a_{12}a_{34}-a_{14}a_{23})(a_{14}a_{23}-
a_{13}a_{24})a_{14}^m ,\, a_{12}a_{14}^m ,\,
a_{12}(a_{12}a_{34}-a_{13}a_{24}+a_{14}a_{23})a_{14}^m ,\,
a_{24}a_{14}^m ,\,
a_{24}(a_{12}a_{34}-a_{13}a_{24}+a_{14}a_{23})a_{14}^m ,\,
a_{34}a_{14}^m ,\, a_{34}(a_{13}a_{24}- a_{14}a_{23})a_{14}^m ,\,
a_{34}a_{24}a_{14}^m ,\, a_{34}a_{12}a_{14}^m \,|\, m=0,...,\ell
-2\}$ and the decomposition reads \bea (\ell -2,1,0,1)\rightarrow
&& \oplus [1,\ell -m-2,m]_{\ell -2m+2}
\oplus [1,\ell -m-2,m]_{\ell -2m-2}\nonumber\\
&& \oplus [1,\ell -m-2,m]_{\ell -2m-6}
\oplus [0,\ell -m-2,m]_{\ell -2m}\nonumber\\
&& \oplus [0,\ell -m-2,m]_{\ell -2m-4}
\oplus [0,\ell -m-1,m+1]_{\ell -2m}\nonumber\\
&&\oplus [0,\ell -m-1,m+1]_{\ell -2m-4}
\oplus [2,\ell -m-2,m]_{\ell -2m}\nonumber\\
&&\oplus [2,\ell -m-2,m]_{\ell -2m-4}
\oplus [1,\ell -m-1,m+1]_{\ell -2m-2}\nonumber\\
&& \oplus [1,\ell -m-2,m]_{\ell -2m-2}
\eea

The relevant $SO(8)\rightarrow SO(6)\times SO(2)$ decomposition is
given by the embedding (\ref{8v}), (\ref{8s}), (\ref{8c}).  In
particular this implies \bea
{\bf 35_v}(2,0,0,0) \rightarrow && {\bf 15}_0[0,1,1] + {\bf 10}_{+2}[0,2,0] + {\bf 10^*}_{-2}[0,0,2]\nonumber\\
{\bf 35_c}(0,0,2,0) \rightarrow && {\bf 15}_0[0,1,1] + {\bf
10^*}_{+2}[0,0,2] + {\bf 10}_{-2}[0,2,0]\nonumber\\
{\bf 35_s}(0,0,0,2)\rightarrow && {\bf 20'}_0[2,0,0] + {\bf 6}_{+2}[1,0,0] + {\bf 6}_{-2}[1,0,0]+\nonumber\\
&& {\bf 1}_{0}[0,0,0] + {\bf 1}_{+4}[0,0,0]+ {\bf 1}_{-4}[0,0,0]
\eea that are necessary to analyze the spectrum of scalars.

The zero charge spectrum \ie the states which constitute the KK
spectrum of Type IIA supergravity on $\mC\mP^3$ can be easily
identified in the above decompositions. For completeness and
comparison with the original literature \cite{Nilsson:1984bj}, we
collect the relevant formulae in an Appendix.

\subsection{A closer look at the KK spectrum}

As already observed, the $\mZ_k$ orbifold projection from $S^7$ to
$S^7/\mZ_k \approx \mC\mP^3 \ltimes S^1$ cannot be thought of as
spontaneous (super)symmetry breaking. `Untwisted' states that are
projected out do not simply become `massive' but are rather
eliminated from the spectrum. In particular in the large $k$ limit
only $SO(2)$ singlets survive. It is amusing to observe that only
states with $\ell$ even on $S^7$ give rise to neutral states. This
suggests that the parent theory could be either a compactification
on $S^7$ or on $\mR\mP^7=S^7/\mZ_2$. Indeed both lead to $SO(8)$ gauged
supergravity corresponding to the `massless' multiplet \be
\{g_{\mu\nu}, 8 \psi_\mu, 28 A_\mu, 56 \lambda, 35^+ + 35^-
\varphi \} \ee

Massless scalars, corresponding to marginal operators with $\Delta
=3$ on the boundary, only appear in higher KK multiplets, \ie in
the ${\bf 840'}=(2,0,0,2)$ and ${\bf 1386}=(6,0,0,0)$. None of
these can play the role of St\"uckelberg field for the 12 coset
vectors in the ${\bf 6}_{+2} + {\bf 6}_{-2}$ of $SO(8) /
SO(6)\times SO(2)$.

Indeed, using the group theory techniques described in Section
\ref{PolyRep} or otherwise, the decomposition of ${\bf
840'}=(2,0,2,0)$ under $SO(8) \rightarrow SO(6)\times SO(2)$ reads
\bea {\bf 840_{vc}}(2,0,2,0) \rightarrow && {\bf 84}_{+4}[0,2,2] +
{\bf 70}_{+2}[0,3,1]+{\bf 70}_{+2}[0,1,3]+ {\bf 64}_{+2}
[1,1,1]\nonumber\\
&& +{\bf 84}_{0}[0,2,2]+{\bf 45}_{0}[1,2,0]+{\bf 45}_{0}[1,0,2]\\
&& +{\bf 35}_{0}[0,4,0]+{\bf 35}_{0}[0,0,4]+{\bf 20}'_{0}[2,0,0]\nonumber\\
&& +{\bf 84}_{-4}[0,2,2] + {\bf 70}_{-2}[0,3,1]+{\bf
70}_{-2}[0,1,3]+ {\bf 64}_{-2} [1,1,1]\nonumber \eea This means
that the massless scalars in the ${\bf 840_{vc}}(2,0,2,0)$ cannot
account for the `needed' St\"uckelberg fields in the ${\bf
6}_{+2}+ {\bf 6}_{-2}$. Yet one can recognize massless scalars
neutral under $SO(2)$ that survive in $k\rightarrow \infty$ limit
and transform non-trivially under $SO(6)$. Turning them on in the
bulk, \eg in domain-wall solutions, should trigger RG flows to
theories with lower supersymmetry on the boundary.

The same applies to the other massless scalars in the ${\bf
1386}(6,0,0,0)$, the totally symmetric product of 6 ${\bf 8_v}
\rightarrow {\bf 4}_{+1}+{\bf 4^*}_{-1}$. The relevant
decomposition reads \bea {\bf 1386}(6,0,0,0) \rightarrow
&& {\bf 84}_{+6}[0,6,0] + {\bf 189}_{+4}[0,5,1]+ {\bf 270}_{+2}[0,4,2]
\nonumber\\
&& +{\bf 300}_{0}[0,3,3]\nonumber\\
&& +{\bf 84}_{-6}[0,0,6]+ {\bf 189}_{-4}[0,1,5]+{\bf 270}_{-2}[0,2,4]
\eea
Once again there are no ${\bf 6}_{+2}+
{\bf 6}_{-2}$. In this case, `neutral' fields appear in the $ {\bf
300}$ representation of $SO(6)$.

In the KK spectrum, neutral (wrt to $SO(2)$) singlets (of $SO(6)$)
appear in the decomposition of ${\bf 35}_{s}$ parity odd scalars
$0_2^-$ with $M^2L_{AdS}^2 = 10$ that reads \bea {\bf
35_s}(0,0,0,2)\rightarrow && {\bf 20'}_0[2,0,0] + {\bf
6}_{+2}[1,0,0]
+ {\bf 6}_{-2}[1,0,0]\nonumber\\
&& +{\bf 1}_{0}[0,0,0] + {\bf 1}_{+4}[0,0,0]+ {\bf 1}_{-4}[0,0,0]
\eea They correspond to boundary operators with dimension $\Delta
= 5$. The only other neutral singlets arise from the $SO(8)$
singlet parity even scalar with $M^2L_{AdS}^2 = 18$, \ie $\Delta
=6$. Neither ones belongs in the supergravity
multiplet\footnote{After gauging $SO(8)$, the 70 scalars give rise
to ${\bf 35_v}(2,0,0,0)$ and ${\bf 35_c}(0,0,2,0)$ which in turn
decompose into ${\bf 35_v}(2,0,0,0) \rightarrow {\bf 15}_0[0,1,1]
+ {\bf 10}_{+2}[0,2,0] + {\bf 10^*}_{-2}[0,0,2]$ and $ {\bf
35_c}(0,0,2,0) \rightarrow {\bf 15}_0[0,1,1] + {\bf
10^*}_{+2}[0,0,2] + {\bf 10}_{-2}[0,2,0]$.}. They correspond to
the `stabilized' complexified K\"ahler deformation $\cJ + i B$ and
as such couple to the Type IIA world-sheet instanton recently
identified in \cite{Cagnazzo:2009zh}. Indeed the bosonic action
schematically reads $S_{wsi} = \int \cJ + i B = L^2/\ap$ since
$B=0$ in the ABJM model, while $B=l/k$ with $l=1,...,k-1$ for the
ABJ model involving fractional M2-branes. Effects induced by
world-sheet instantons in Type IIA on $\mC\mP^3$ should be dual to
the non-perturbative corrections discussed in
\cite{Hosomichi:2008ip}. It may be worth to observe that in
`ungauged' $\cN=6$ supergravity, arising from freely acting
asymmetric orbifolds of Type II superstrings on tori, world-sheet
and other asymmetric brane instantons \cite{Bianchi:2009mu,
Bianchi:2008cj} should correct $\cR^4$ terms very much as in their
parents with $\cN=8$ local supersymmetry.

Other non-perturbative effects are induced by E5-brane instantons
that should mediate the process of annihilation of $k$ D0-branes
into $N$ D4-branes wrapping $CP^2$
\cite{Aharony:2008gk,Aharony:2008ug}. In order to determine the
action of such an instanton it is worth recalling that the
pseudo-scalar mode $B_2 = \beta(x)J_2(y)$ is eaten by the vector
field $A_\mu^H = kA_\mu^{D4} - N A_\mu^{D0}$ that becomes massive.
The complete E5-brane instanton action should be $S_{E5} =
L^6/g_s^2(\ap)^3 + i \beta$ that indeed shifts under $U(1)_{H}$
gauge transformations and as such can compensate for the `charge'
violation in the above process as in similar cases with unoriented
D-brane instantons \cite{Bianchi:2009ij}.

\section{Singleton, partition functions and Higher Spins}

In this section, we would like to discuss the higher spin (HS)
extension of $\cN = 6$ gauged supergravity.
Higher spin extensions of various supergravity theories in $AdS_4$
have been studied in \cite{Sezgin:1998gg, Sezgin:2002rt,
Engquist:2002vr} but to the best of our knowledge the case of $\cN
= 6$  has been overlooked.

Let us start by briefly recalling some basic features of higher
spin theories in $AdS_4$\footnote{See \eg \cite{Bianchi:2004ww,
Bianchi:2004xi, Bekaert:2005vh, Francia:2006hp} for recent
reviewes of both Vasiliev's and geometric approaches.}.
 In the non supersymmetric case the
HS algebra represents an extension of the conformal group
$SO(3,2)$ that admits two {\it singleton} representations
$\cD(1/2,0)$ (free boson) and $\cD(1,1/2)$ (free fermion). The two
labels denote conformal dimension $\Delta$ and spin $s$. Indeed
the maximal compact subgroup of $SO(3,2)$ is $SO(3)\times SO(2)
\approx SU(2)\times U(1)$ while `Lorentz' transformations and
dilatations commute and generate $SO(2,1)\times SO(1,1) \subset
SO(3,2)$. We will continue and call $\Delta$ the dimension and $s$
or $j$ spin. In `radial' quantization the `Hamiltonian' $\cH$ has
eigenvalues $\Delta$.

For later use let us collect here the partition functions of the
two singletons that take into account their conformal descendants
\ie non vanishing derivatives. For free bosons such that $\de^2
X=0$ one has \be \cZ_B(q) = Tr q^{2\cH} = {q - q^5 \over
(1-q^2)^3} ={q + q^3 \over (1-q^2)^2}\ee For free fermions
$\slash{\,\de} \Psi=0$ one has \be \cZ_F(q) = Tr q^{2\cH}= 2{q^2 -
q^4 \over (1-q^2)^3} = 2{q^2\over (1-q^2)^2} \ee Combining
$n_b=8_v$ free bosons and $n_f=8_c$ free fermions one finds the
singleton representation of $Osp(8|4) \supset SO(8)\times
SO(3,2)$, whose Witten index reads \bea \cZ_{\tiny\yng(1)}(q)=
Tr(-)^F q^{2\cH} = 8_v \cZ _B(q) - 8_c \cZ_F(q)\eea One can also
keep track of the spin of the states in the spectrum by including
a chemical potential $y=e^{i\alpha}$ ($y^{J_3} =e^{i\alpha J_3}$)
and find \bea \cZ _B(q, \alpha)=\frac{q(1-q^4)}{(1-q^2) (1- e
^{i\alpha} q^2) (1- e ^{-i\alpha} q^2)}= \frac{q(1+q^2)}{(1-2 q^2
\cos\alpha +q^4)} \eea \bea \cZ_F(q, \alpha)=\frac{q^2 (1 -
q^2)\chi _{1\over 2}(\alpha)}{(1-q^2) (1-2 q^2 \cos\alpha + q^4)}
\eea where \bea \chi _{1\over 2}(\alpha) = 2 \cos \frac{\alpha}{2}
= tr_{1/2} e^{i\alpha J_3} \eea is the character of the
fundamental representation of the `Lorentz' group $SU(2)$.

Before switching to higher spins, notice that $\mZ_k$ acts on the
singleton simply as \be 8_v\rightarrow 4\omega + 4^* \bar{\omega}
\qquad 8_c\rightarrow 4 \bar{\omega} + 4^* \omega \qquad
8_s\rightarrow 6 +  {\omega}^2 + \bar\omega^2 \ee with
$\omega=e^{2\pi i/k}$ playing the role of chemical potential or
rather fugacity for the $SO(2)\approx U(1)_B$ charge $Q$ commuting
with $SO(6)$ R-symmetry. One can introduce another three chemical
potentials $\beta_i$ or fugacities $x_i = e^{i\beta_i}$ in order
to keep track of the three Cartan's of $SO(6)\approx SU(4)$. We
refrain from doing so here.

\subsection{Doubleton and higher spin gauge fields}

{\it Doubleton} representations can be obtained as tensor products
of two singletons \cite{Ferrara:1997dh, Sundborg:1999ue,
HaggiMani:2000ru}. \be \cD(1/2,0)\otimes \cD(1/2,0) =
\oplus_{s=0}^\infty \cD(\Delta=s+1,s) \ee or \be \cD(1,
1/2)\otimes \cD(1,1/2) = \cD(\Delta=2,s=0) + \oplus_{s\neq
0}^\infty \cD(\Delta=s+1,s) \ee A consistent truncation, giving
rise to minimal HS theories with even spins only, stems from
restricting to symmetric tensors for bosons \be [\cD(1/2,0)\otimes
\cD(1/2,0)]_S = \oplus_{k=0}^\infty \cD(\Delta=2k+1,s=2k) \ee or
anti-symmetric for fermions \be [\cD(1, 1/2)\otimes \cD(1,1/2)]_A
= \cD(\Delta=2,s=0) + \oplus_{k\neq 0}^\infty
\cD(\Delta=2k+1,s=2k) \ee Odd spin states appear in the product
with opposite symmetry \be [\cD(1/2,0)\otimes \cD(1/2,0)]_A =
\oplus_{k=0}^\infty \cD(\Delta=2k+2,s=2k+1) \ee for bosons and \be
[\cD(1, 1/2)\otimes \cD(1,1/2)]_S = \oplus_{k=0}^\infty
\cD(\Delta=2k+2,s=2k+1) \ee for fermions. Generators of the HS
symmetry algebra can be realized as polynomials of bosonic
oscillators $y_\a, y_{\dot\a}=(y_\a)^\dagger$ satisfying $[y_\a,
y_\b] = i\varepsilon_{\a\b}$ and $[y_{\dot\a}, y_{\dot\b}] =
i\varepsilon_{\dot\a\dot\b}$.

The supersymmetric extensions require the introduction of
fermionic oscillators $\xi^i$ with $i=1, ..., \cN$, satisfying
$\{\xi^i, \xi^j\} = \delta^{ij}$. The resulting HS superalgebra
denoted by $shs^E(\cN|4)$ contains $Osp(\cN|4)$ whose bosonic
generators span $SO(3,2)\cong Sp(4,R)$ (conformal group) and
$SO(\cN)$ R-symmetry \cite{Sezgin:1998gg, Sezgin:2002rt,
Engquist:2002vr}.

In particular for $\cN =8$, with $SO(8)$ R-symmetry, $Osp(8|4)$ is
the maximal finite dimensional subalgebra of the HS gauge algebra
$shs^E(8|4)$, which is a Lie superalgebra. The relevant
super-singleton consists in\footnote{Different conventions for the
$SO(8)$ representations of bosons and fermions appear in the
literature which are related to the present one, chosen for
compatibility with our previous analysis, by $SO(8)$ triality.}
\be \widehat\cD_{\cN=8} = \cD(1/2,0;{\bf 8}_v) \oplus
\cD(1,1/2;{\bf 8}_c) \ee The (graded) symmetric product of two
singletons $[\widehat\cD_{\cN=8} \otimes
\widehat\cD_{\cN=8}]_{\hat{S}}$ yields \bea &&\{[\cD(1/2,0;{\bf
8}_v) \oplus \cD(1,1/2;{\bf 8}_c)] \otimes [\cD(1/2,0;{\bf 8}_v)
\oplus
\cD(1,1/2;{\bf 8}_c)]\}_{\hat{S}} = \nn \\
&&\cD(1,0;{\bf 1} + {\bf 35}_v) \oplus \cD(2,0;{\bf 1} + {\bf
35}_c) \oplus_k \cD(k+{3\over 2},k+{1\over 2}; {\bf 8}_s + {\bf
56}_s)\nn \\&&\oplus_{k\neq 0} \cD(2k+1,2k; {\bf 1} + {\bf 35}_v +
{\bf 1} + {\bf 35}_c) \oplus_k \cD(2k+2,2k+1; {\bf 28} + {\bf 28})
\eea

It is reassuring to recognize above the `massless' states of $\cN
= 8$ gauged supergravity on $AdS_4$. The remaining states with
spin $s\le 2$ belong to the `short' Konishi multiplet and a
`semishort' multiplet with spin ranging from 2 to 6
\cite{Bianchi:1999ge, Bianchi:2000hn, Bianchi:2001cm }. Holography
allows to relate AdS compactifications of supergravity and
superstring theories to singleton field theories on the 3-d
boundary. As a first step, these field theories can be constructed
on the boundary of AdS as free superconformal theories. A
remarkable property of singletons is that the symmetric product of
two super-singletons gives an infinite tower of massless higher
spin states. In the limit $\lambda \rightarrow 0$, all higher spin
states become massless. After turning on interactions, a
pantagruelic Higgs mechanism, named {\it Grande Bouffe} in
\cite{Bianchi:2003wx, Beisert:2003te, Beisert:2004di,
Bianchi:2005ze}, takes place. All but a handful of HS gauge fields
become massive after `eating' lowest spin states. The boundary
counterpart of this phenomenon is the appearance of anomalous
dimensions for HS currents and their superpartners. One should
keep in mind that genuinely massive states are already present in
the spectrum at $\lambda \rightarrow 0$ and arise in the product
of three and more singletons.

Interacting theories for massless HS gauge fields, thus only
describing the doubleton, have been proposed by Vasiliev
\cite{Bekaert:2005vh} that capture some aspects of the holographic
correspondence in the extremely stringy (high AdS curvature)
regime. Only vague glimpses of an interacting theory incorporating
the {\it Grande Bouffe} have been offered so far
\cite{Bianchi:2003wx, Beisert:2003te, Beisert:2004di,
Bianchi:2005ze}.

Barring these subtle issues, let us discuss how to perform a
$\mZ_k$ projection of the spectrum giving rise to an $\cN =6$ HS
supergravity in $AdS_4$. In the limit $k\rightarrow \infty$ only
$SO(2)$ singlets survive \bea &&\{[\cD(1/2,0;{\bf 8}_v) \oplus
\cD(1,1/2;{\bf 8}_c)]
\}^{\otimes 2_{\hat{S}}}_{SO(2) singlets} = \nn \\
&&\cD(1,0;{\bf 1} + {\bf 15}) \oplus \cD(2,0; {\bf 1} + {\bf 15})
\oplus_k \cD(k+{3\over 2},k+{1\over 2}; {\bf 6} + {\bf 6} + {\bf
10} + {\bf 10}^*)\nn
\\&&\oplus_{s\neq 0} \cD(s+1,s; {\bf 1} + {\bf 15} + {\bf 1} +
{\bf 15})  \eea where indicated in bold-face are the surviving
representations of the $SO(6)$ R-symmetry. Candidate bosonic HS
operators on the boundary in the ${\bf 1} + {\bf 15}$ of $SO(6)$
are \be \cJ_{\mu_1...\mu_s}{}^i{}_j = X^i \de_{\mu_1}\de_{\mu_2}
... \de_{\mu_s} \bar{X}_j +  \bar{\Psi}^i \gamma_{\mu_1} \de_{\mu_2} ...
\de_{\mu_s} \Psi_j + ... \ee where dots stand for
symmetrization and subtraction of the traces and the coefficients
of the linear combination are to be chosen appropriately.

At finite $k$ and $\lambda$, states with $SO(2)$ charges $Q = k n$
survive. One can exploit orbifold technique to deduce the `free'
spectrum\footnote{Although $k$ is finite, one can take $k>>N$, so
that $\lambda<< 1$, in order to identify states that eventually
become massive.}.

The partition function or rather Witten index for the
super-singleton of $OSp(8|4)$ reads: \bea \cZ_{\tiny\yng(1)}
=\frac{8q}{(1+q)^2} \eea the $\mZ_k$ projection reads \be
\cZ^{\mZ_k}_{\tiny\yng(1)} = {1\over k} \sum_{r=0}^{k-1}
\cZ^{(r)}_{\tiny\yng(1)} \ee where

\be \cZ^{(r)}_{\tiny\yng(1)} =\frac{(4 \omega^r + 4\bar\omega^r)
q}{(1+q)^2} \ee with $\omega = e^{2\pi i/k}$. Clearly
$\cZ^{\mZ_k}_{\tiny\yng(1)} = 0$ since $\Sigma_{r=0}^{k-1}
\omega^r = 0$.

A non-trivial spectrum arises from the doubleton partition
function. Prior to the $\mZ_k$ projection one has \bea
\cZ_{\tiny\yng(2)}=\frac{1}{2}(\cZ_{\tiny\yng(1)} ^2(q) +
\cZ_{\tiny\yng(1)}(q^2))=4q^2 (8 (1+q)^{-4}+(1+q^2)^{-2}) \eea for
the (graded) symmetric doubleton, giving rise to precisely the
spectrum of $hs(8|4)$ discussed above.

Performing the $\mZ_k$ projection on the symmetric doubleton one
finds \bea &&\cZ^{\mZ_k}_{\tiny\yng(2)}=\frac{1}{2k}\sum_r
(\cZ^{(r)}_{\tiny\yng(1)}(q,\omega)^2 +
\cZ^{(r)}_{\tiny\yng(1)}(q^2,\omega^2)) \nn \\ && =
4q^2\left[4\left( 1 + \sum_r{\omega^{2r} + \bar\omega^{2r} \over
2k}\right)(1+q)^{-4}+\sum_r{\omega^{2r} + \bar\omega^{2r} \over
2k}(1+q^2)^{-2}\right] \eea for the (graded) symmetric doubleton,
giving rise to precisely the `massless' HS gauge fields of
$hs(6|4)$ for $k\neq 2$ and $hs(8|4)$ for $k = 1, 2$, as expected
$Z_{HS}=Z_{\tiny\yng(2)}$! Indeed \bea && \cZ_{HS} = \frac{36
(q^2 +q^4) + 72 \sum _{s=2k\neq 0}F_s(q) + 56 \sum _{s=2k+1}F_s(q)
- 64 \sum _{s=k+ \frac{1} {2}}F_s(q)}{(1-q^2)^3}\nonumber\\ \eea
with $F_s(q) = (2s+1) q^{2(s+1)} - (2s-1) q^{2(s+1)+2}$ taking
into account the presence of null descendants for conserved spin
$s$ currents of dimension $\Delta = s +1$. The relevant characters
read \bea \cX _s ^{\Delta =s+1} = \frac{q ^{2 \Delta}(2s+1) -
q^{2(\Delta +1)} (2s-1) }{(1 - q^2)^3} =
\frac{q ^{2 \Delta} [\chi _s(\alpha) - q^2 \chi _{s-1}(\alpha)
]}{(1-q^2) (1-2 q^2 \cos\alpha + q^4)} \eea up to some $SO(8)$
multiplicity $d_{(\ell ,...)}^{SO(8)}$.

The situation is summarized in the following Tables, where $s$
denotes spin and $h$ the `string' level.
\begin{table}
\begin{center}
 \begin{tabular}{ | l | l | l | l | l |}
  \hline
    $s \backslash h$ & 0 & 1 & 2 & 3 \\ \hline
    0 & 70 & 1+1 &  & \\ \hline
    $1\over 2$ & 56 & 8 &  &\\ \hline
    1 & 28 & 28 &  &\\ \hline
    $3\over 2$ & 8 & 56 &  &\\ \hline
    2 & 1 & 70 & 1 &\\ \hline
    $5\over 2$ &  & 56 & 8 &\\ \hline
    3 &  & 28 & 28 &\\ \hline
    $7\over 2$ &  & 8 & 56 &\\ \hline
    4 &  & 1 & 70 & 1 \\ \hline
    ... &  &  & ... & ...\\ \hline
    \end{tabular}
    \caption{$\cN=8$ $hs(8|4)\supset Osp(8|4)$}
\end{center}
\end{table}

The decomposition into charged sectors reads \bea
\cZ_{\tiny\yng(2)}=&&\frac{1}{(1-q^2)(1-2q^2\cos \alpha +q^4)}\, \{ \left[10 \left(\omega
^2+\omega_c^2\right)+16\right]\left(q^2+q^4\right)\chi_0(y)\nonumber\\
&&+\sum_{j\in 1,3,...}\left[12\left(\omega
^2+\omega_c^2\right)+32\right]
\left[\chi_j(y) q^{2 (j+1)}-\chi_{j-1}(y) q^{2 (j+1)+2}\right] \nonumber\\
&&+\sum_{j\in 2,4,...}\left[20\left(\omega
^2+\omega_c^2\right)+32\right] \left(\chi_{j}(y) q^{2
(j+1)}-\chi_{j-1}(y) q^{2
(j+1)+2}\right)\nonumber\\
&&-\sum_{j\in 1/2,3/2,...}16(\omega +\omega_c)^2\left(\chi_{j}(y)
q^{2 (j+1)}-\chi_{j-1}(y) q^{2 (j+1)+2}\right)\}. \eea
\begin{table}
\begin{center}
 \begin{tabular}{ | l | l | l | l | l | }
    \hline
    $s\backslash h$ & 0 & 1 & 2 & 3 \\ \hline
    0 & 15+15 & 1+1 &  &\\ \hline
    $1\over 2$ & $10+10^*+6$ & 6 &  &\\ \hline
    1 & 15+1 & 15+1 & & \\ \hline
    $3\over 2$ & 6 & $10+10^*+6$ &  &\\ \hline
    2 & 1 & 15+15 & 1 &\\ \hline
    $5\over 2$ &  & $10+10^*+6$ & 6 & \\ \hline
    3 &  & 15+1 & 15+1 & \\ \hline
    $7\over 2$ &  & 6 & $10+10^*+6$ & \\ \hline
    4 &  & 1 & 15+15 & 1  \\ \hline
    $9\over 2$ &  &  & $10+10^*+6$ & 6  \\ \hline
    5 &  &  & 15+1 & 15+1 \\ \hline
    $11\over 2$ &  &  & 6 & $10+10^*+6$ \\ \hline
    6 &  &  & 1+1 & 15+15  \\ \hline
    ... &  &  & ... & ...  \\ \hline
    \end{tabular}
\caption{$SO(2)$ neutral HS for $\cN=6$: $hs(6|4)\supset Osp(6|4)$}
\end{center}
\end{table}

\begin{table}
\begin{center}
 \begin{tabular}{ | l | l | l | l | l |}
    \hline
    $s\backslash h$ & 0 & 1 & 2 &\\ \hline
    0 & $(10+10^*)_{\pm 2}$ &  &  &\\ \hline
    $1\over 2$ & $15_{\pm 2}$ & $1_{\pm 2}$ &  &\\ \hline
    1 & $6_{\pm 2}$ & $6_{\pm 2}$ &  &\\ \hline
    $3\over 2$ & $1_{\pm 2}$ & $15_{\pm 2}$ &  &\\ \hline
    2 &  & $(10+10^*)_{\pm 2}$ & &\\ \hline
    $5\over 2$ &  & $15_{\pm 2}$ & $1_{\pm 2}$ &\\ \hline
    3 &  & $6_{\pm 2}$ & $6_{\pm 2}$ &\\ \hline
    $7\over 2$ &  & $1_{\pm 2}$ & $15_{\pm 2}$ &\\ \hline
    4 &  &  & $(10+10^*)_{\pm 2}$ &\\ \hline
    ... &  &  & ...& ... \\ \hline
    \end{tabular}
    \caption{Charged HS for $\cN=6$: $hs(8|4)/hs(6|4)\supset Osp(8|4)/Osp(6|4)$}
\end{center}
\end{table}

\subsection{Tripletons and higher $n$-pletons}

For higher multipletons one has to resort to Polya theory
\cite{Bianchi:2003wx, Beisert:2003te, Beisert:2004di,
Bianchi:2005ze}. Consider a set of `words' $A,B,...$ of $n$
`letters' chosen within the alphabet $\{a_i\}$ with $i = 1,... p$.
When $p\rightarrow \infty$, let us denote by $\cZ_1(q)$ the single
letter `partition function'. Let also $G$ be a group action
defining the equivalence relation $A\sim B$ for $A = gB$ with $g$
an element of $G\subset S_n$. Elements $g \in S_n$ can be divided
into conjugacy classes $[g] = (1)^{b_1}... (n)^{b_n}$, according
to the numbers $\{b_k(g)\}$ of cycles of length $k$. Polya theorem
states that the set of inequivalent words are generated by the
formula \be \cZ_n^G = {1\over |G|} \sum_{g\in G}
\prod_{k=1}^n\cZ_1(q^k)^{b_k(g)} \ee

In particular, for the cyclic group $G = Z_n$, conjugacy classes
are $[g] = (d)^{n/d}$ for each divisor $d$ of $n$. The number of
elements in a given conjugacy class labelled by $d$ is given by
Eulers totient function $\cE(d)$, equal to the number of integers
relatively prime to $d$. For $d = 1$ one defines $\cE(1)=1$. \be
\cZ_n^{Z_n} = {1\over n} \sum_{d | n} \cE(d)\cZ_1(q^d)^{n/d} \ee
For the full symmetric group one has \be \cZ_n^{S_n} = {1\over n!}
\sum_{n_r: \sum_r r n_r = n} {n! \over \prod_r r^{n_r} n_r!}
\prod_r \cZ_1(q^{r})^{n_r} \ee

Let us consider the product of three singletons. \bea
\cZ_{\tiny\yng(1)}^3 = \cZ_{\tiny\yng(1) \times \tiny\yng(1)
\times \tiny\yng(1)} \rightarrow
\cZ_{\tiny\yng(3)}+\cZ_{\tiny\yng(1,1,1)}+\cZ_{\tiny\yng(2,1)}
\eea There are thus three kinds of tri-pletons.

The totally symmetric tripleton is coded in the partition function
\bea && \cZ_{\tiny\yng(3)} = \frac{1}{6}(\cZ_{\tiny\yng(1)}^3(u) +
3 \cZ_{\tiny\yng(1)}(u)\cZ_{\tiny\yng(1)}(u^2) + 2
\cZ_{\tiny\yng(1)}(u^3))\eea where $u$ collectively denotes the
`fugacities' $q,y=e^{i\alpha},\omega\approx t, ...$.

For the cyclic tripleton one has \bea && \cZ_{cycl} =
\cZ_{\tiny\yng(3)} + \cZ_{\tiny\yng(1,1,1)} =
\frac{1}{3}(\cZ_{\tiny\yng(1)}^3(u) + 2\cZ_{\tiny\yng(1)}
(u^3))\eea

For totally anti-symmetric tripletons one finds \bea &&
\cZ_{\tiny\yng(1,1,1)} = \cZ _{cycl}- \cZ_{\tiny\yng(3)} =
\frac{1}{6}(\cZ_{\tiny\yng(1)}^3(u) + 2\cZ_{\tiny\yng(1)} (u^3) -
3\cZ_{\tiny\yng(1)}(u) \cZ_{\tiny\yng(1)}(u^2)\eea
while for mixed symmetry, incompatible with the cyclicity of the
trace, one eventually finds \bea &&\cZ_{\tiny\yng(2,1)} =
\cZ_{\tiny\yng(1)}^3 (u) - \frac{1}{3} \cZ_{\tiny\yng(1)}^3 (u) -
\frac{2}{3}\cZ_{\tiny\yng(1)} (u^3) = \frac{2}{3}
(\cZ_{\tiny\yng(1)}^3 (u) - \cZ_{\tiny\yng(1)} (u^3)) \eea

Recalling the singleton partition function \bea
\cZ_{\tiny\yng(1)}(q, \alpha , \omega) = &&\frac{(4 \omega + 4^*
\bar{\omega}) q (1+q^2)}{(1-2 q^2 \cos\alpha + q^4)} -
\frac{(4 \bar{\omega} + 4^* \omega) q^2 \chi _{\frac{1}{2}} (\alpha)}{(1-2 q^2 \cos\alpha + q^4)}\nonumber\\
&& =\frac{4 (\omega + \bar{\omega})q}{(1-2 q^2 \cos\alpha + q^4)}
[1 + q^2 - \chi _{\frac{1}{2}} (\alpha) q] \eea where $\omega =
e^{2\pi i/k}$ and $\chi _{\frac{1}{2}} (\alpha)= tr_{1/2} \exp
(i\alpha J_3)$, one eventually finds \bea \cZ_{\tiny\yng(3)} =
&&\frac{1}{6}\left ( \frac{4^3 (\omega + \bar{\omega})^3 q^3 (1 +
q^2 -q \chi _{\frac{1}{2}}
(\alpha) )^3}{(1-2 q^2 \cos\alpha + q^4)^3} + \right .\nonumber\\
&& \left .3 \frac{4 (\omega + \bar{\omega}) q 4 (\omega ^2 +
\bar{\omega} ^2) q^2 (1 + q^2 -q\chi _{\frac{1}{2}} (\alpha))
(1 + q^4 - q^2\chi _{\frac{1}{2}} (2\alpha))}{(1-2 q^2 \cos\alpha + q^4) (1-2 q^4 \cos\alpha + q^8)} + \right .\nonumber\\
&& \left .2 \frac{4 (\omega ^3+ \bar{\omega}^3) q^3 (1 + q^6 -
q^3\chi _{\frac{1}{2}} (3 \alpha))}{(1-2 q^6 \cos\alpha + q^{12})}
\right ) \eea
for the totally symmetric tripleton.
Let us analyze the spectrum arising in this case. Except for the 1/2 BPS states, we will consider later
on, only `massive' representations above the unitary bound, whose
characters read \bea && \cX _s^{\Delta \neq s+1} = \frac{q ^{2
\Delta} \chi _s(\alpha)}{(1-q^2) (1-2 q^2 \cos\alpha + q^4)}
\rightarrow_{\alpha\rightarrow 0}\frac{q ^{2 \Delta}(2s+1)}{(1 -
q^2)^3} \eea appear in the decomposition \bea
\cZ_{\tiny\yng(3)}(q, \alpha , \omega) =\sum _{s, \Delta , Q}
c(s,\Delta, Q)\frac{q ^{2 \Delta} \chi _s(\alpha) \omega ^Q}
{(1-q^2) (1-2 q^2 \cos\alpha + q^4)} \eea Indeed it is easy to see
that no current like (twist $\tau =1$) fields appear beyond the
double-ton, since the twist \be \tau = \Delta -s = {n_X\over 2} +
n_{\de} + n_\Psi - n_{\de} - {n_\Psi\over 2} = {n_X\over 2} +
{n_\Psi\over 2} >1 \ee whenever $n_X + n_\Psi > 2$.

Using orthogonality of the $SU(2)$ characters \bea
\frac{1}{\pi}\int _0^{2\pi} \chi _s(\alpha)  \chi
_{s^\prime}(\alpha) \sin ^2\frac{\alpha}{2} d\alpha =\delta
_{2s+1,2s^\prime +1} \eea

one can decompose the partition function according to \bea
\sum_{Q,\Delta} \frac{c(s,\Delta, Q) \omega ^Q q ^{2
\Delta}}{(1-q^2)}=\frac{1}{\pi}\int _0^{2\pi}(1-2 q^2 \cos\alpha +
q^4) \sin ^2\frac{\alpha}{2} \chi _s(\alpha)\cZ_{\tiny\yng(3)}(q,
\alpha , \omega)  d\alpha \nn \\\eea

It is clear that only states with charge $Q=\pm 3, \pm 1$ are
present in the tri-pleton spectrum. Setting $y=e^{i\alpha}$, for
states with $Q=\pm1$ one finds \bea
\cZ_{\tiny\yng(3)}^{Q=\pm 1}&=&\sum_{k=0}^{\infty}\left[(40+256 k) q^{4 k+3}+(104+256 k) q^{4 k+5}\right] \chi_{2 k}
(y)\nonumber\\
&-&\left[(104+256 k) q^{4 k+4}+(152+256 k) q^{4 k+6}\right] \chi_{2 k+\frac{1}{2}}(y)\nonumber\\
&+&\left[(152+256 k) q^{4 k+5}+(216+256 k) q^{4 k+7}\right] \chi_{2 k+1}(y)\nonumber\\
&-&\left[(216+256 k) q^{4 k+6}+(296+256 k) q^{4 k+8}\right]
\chi_{2 k+\frac{3}{2}}(y) \eea these states are always projected
out by $\mZ_k$ since $\pm 1\neq n k$. For states with $Q=\pm3$ one
finds instead \bea
\cZ_{\tiny\yng(3)}^{Q=\pm 3}&=&\sum_{k=0}^{\infty}\left[\left[(20+256 k) q^{12 k+3}+(40+256 k) q^{12 k+5}\right] \chi_{6
k}(y)\right.\nonumber\\
&-&\left[(40+256 k) q^{12 k+4}+(44+256 k) q^{12 k+6}\right] \chi_{6 k+\frac{1}{2}}(y)\nonumber\\
&+&\left[(44+256 k) q^{12 k+5}+(68+256 k) q^{12 k+7}\right] \chi_{6 k+1}(y)\nonumber\\
&-&\left[(68+256 k) q^{12 k+6}+(104+256 k) q^{12 k+8}\right] \chi_{6 k+\frac{3}{2}}(y)\nonumber\\
&+&\left[(104+256 k) q^{12 k+7}+(124 +256 k) q^{12 k+9}\right] \chi_{6 k+2}(y)\nonumber\\
&-&\left[(124+256 k) q^{12 k+8}+(132+256 k) q^{12 k+10}\right] \chi_{6 k+\frac{5}{2}}(y)\nonumber\\
&+&\left[(132+256 k) q^{12 k+9}+(152+256 k) q^{12 k+11}\right] \chi_{6 k+3}(y)\nonumber\\
&-&\left[(152+256 k) q^{12 k+10}+(188+256 k) q^{12 k+12}\right] \chi_{6 k+\frac{7}{2}}(y)\nonumber\\
&+&\left[(188+256 k) q^{12 k+11}+(212+256 k) q^{12 k+13}\right] \chi_{6 k+4}(y)\nonumber\\
&-&\left[(212+256 k) q^{12 k+12}+(216+256 k) q^{12 k+14}\right] \chi_{6 k+\frac{9}{2}}(y)\nonumber\\
&+&\left[(216+256 k) q^{12 k+13}+(236+256 k) q^{12 k+15}\right] \chi_{6 k+5}(y)\nonumber\\
&-&\left. \left[(236+256 k) q^{12 k+14}+(276+256 k) q^{12
k+16}\right] \chi_{6 k+\frac{11}{2}}(y)\right].\,\, \eea These states
survive only for $k=3$, \ie $Z_3$ projection. It is amusing to
observe how the number of representations of given spin $s=6
k+\frac{n}{2}$ grows with $k$ at the rate $256k$ for any $n$. This
is due to the possible distributions of derivatives among three
fields up to symmetries and total derivatives and to the structure
of higher spin supermultiplets \cite{Bianchi:2006ti}.

For higher multi-pletons the analysis is similar. It is clear that
only states with charge $Q=\pm n, \pm (n-2), ...$ are present in
the n-pleton spectrum. In particular $Q=0$ states are only present
when $n$ is even as already observed. We defer a detailed analysis to the future. For
the time being let us only display the partition functions for the
cyclic tetrapleton \bea
\cZ_{4,cycl}=\frac{1}{4}(\cZ_{\tiny\yng(1)}(q)^4 +
\cZ^2_{\tiny\yng(1)}(q^2)+ 2 \cZ_{\tiny\yng(1)}(q^4)) \eea and for the
totally symmetric tetrapleton \bea
\cZ_{\tiny\yng(4)}=\frac{1}{4!}(\cZ_{\tiny\yng(1)} ^4(q) + 6
\cZ^2_{\tiny\yng(1)}(q) \cZ_{\tiny\yng(1)}(q^2)+3 \cZ^2_{\tiny\yng(1)}
(q^2)+8 \cZ_{\tiny\yng(1)}(q^3)\cZ_{\tiny\yng(1)}(q)+6
\cZ_{\tiny\yng(1)}(q^4))\nn\\ \eea

The $\mZ_k$ projection on n-pletons reads \be \cZ_n^{\mZ_k} = {1\over
k} \sum_r \cZ_n^{(r)}(q,\omega^r) \ee and corresponds to keeping
only states with $Q=k n$ \ie integer multiples of $k$.

\subsection{KK excitations}

Let us now focus on the KK excitations, which deserve a separate
treatment. One can indeed write down the single-particle partition
function on $S^7$, decompose it into super-characters and identify
the $SO(2)$ charge sectors, relevant for the subsequent $\mZ_k$
projection \ie compactification on $\mC\mP^3$.

Introducing a chemical potential for the charge $Q$ ($t^Q$), the
super-character of an ultra-short 1/2 BPS representation of
$Osp(8|4)$ reads: \bea &&
\cX^{1/2BPS}_{\ell}(q,t)=\frac{t^{-2-\ell} q^{2+\ell}}{6
\left(1-t^2\right)^5 (1+q)^3}
\left[\ell^3 \left(-1+t^2\right)^2 (-1+q)^3 \right.\nonumber\\
&& \times \left(t^{6+2 \ell} \left(t^2-q\right)^2-\left(-1+t^2 q\right)^2\right)-6 \ell^2 \left(-1+t^2\right) (-1+q)^2
\nonumber\\
&& \times \left(t^{6+2 \ell} \left(t^2-q\right)^2 \left(-3+2 t^2+q\right)+\left(2+t^2 (-3+q)\right) \left(-1+t^2
q\right)^2\right)\nonumber\\
&& +6 t^{6+2 \ell} \left(t^2-q\right)^2 \left(-35+q (35+(-9+q) q)+2 t^4 \left(-5+q^2\right)\right.\nonumber\\
&& \left.+t^2 (35+q (-13+(-7+q) q))\right)
-\left(2 \left(-5+q^2\right)\right. \\
&&\left. +t^4 (-35+q (35+(-9+q) q))+t^2 (35+q (-13+(-7+q) q))\right)\nonumber\\
&& \times 6 \left(-1+t^2 q\right)^2 -\ell (-1+q)\left(t^{6+2 \ell} \left(t^2-q\right)^2 \left(-107+(70-11 q) q \right.
\right.\nonumber\\
&& \left. +t^4 (-47+(-2+q) q)-2 t^2 (-71+q (22+q))\right)
+\left(-1+t^2 q\right)^2 \nonumber\\
&&\left.\left.\times \left(47-(-2+q) q+2 t^2 (-71+q (22+q))+t^4
(107+q (-70+11 q))\right)\right)\right]\nonumber \eea

For $\ell =0$, corresponding to the gauged supergravity multiplet,
there is further shortening (null descendants) due to the presence
of conserved `currents' \ie stress-tensor, $SO(8)$ vector currents
and ${\bf 8}_s$ supercurrents. Taking this into account one finds
the following super-character \bea
\cX^{1/2BPS}_{\ell =0}(q)=\frac{1}{(1-q^2)^3}&& [(10t^2 +15 +10t^{-2})q^2-\nonumber\\
&& 2(15t^2 +10 +6 +10 +15t^{-2})q^3+\nonumber\\
&& (10t^2 +15 +10t^{-2} +3(6t^2+ 15+ 1+ 6t^{-2}))q^4-\nonumber\\
&& 4(t^2+ 6+ t^{-2})q^5 -(6t^2+ 15+ 1- 5+ 6t^{-2})q^6+\nonumber\\
&& 2(t^2+ 6+ t^{-2})q^7 -3q^8] \eea the denominator takes into
account derivatives (descendants). Quite remarkably this formula
coincides with the previous one when $\ell =0$.

After some algebra, putting $t=1$, one finds \be
\cX^{1/2BPS}_{\ell =0}(q) = {q^2 (3 q^3 - 7 q^2 - 7 q +35) \over
(1+q)^3}\ee a factor $(1-q)^2$
 cancels between numerator and denominator
meaning that not only $n_b=n_f$ and the sum with $\Delta^1$
vanishes but also the sum with $\Delta^2$ should vanish. This should
be related to the absence of quantum corrections to the negative vacuum energy,
\ie cosmological constant in the bulk.

The 1/2 BPS partition function is given by \be
\cZ^{\cN=8}_{1/2BPS} = \sum_\ell \cX^{1/2BPS}_\ell = {35 q^2 \over
(1-q^2)^2} \ee The simplicity of the result is due to `miraculous'
cancellations between bosonic and fermionic operators with the
same scaling dimensions in different KK multiplets \ie with
different $\ell$'s. This does not happen in $AdS_5/CFT_4$
holography, whereby (protected) bosonic operator have integer
dimensions and (protected) fermionic operators have half-integer
dimensions \cite{Bianchi:2003wx, Beisert:2003te, Beisert:2004di,
Bianchi:2004ww, Bianchi:2004xi, Bianchi:2005ze, Bianchi:2006ti}.

In order to perform the $\mZ_k$ projection it is useful to decompose
into $SO(2)$ charge sectors according to \be \cZ^{\cN=8
\rightarrow \cN=6}_{1/2BPS}(q,t) = {q^2 [(1+q^6) P_2(t) - (q +
q^5) P_3(t) + (q^2 + q^4) P_4(t) - q^3 P_5(t) ] \over (1-q t)^4
(1-q t^{-1})^4
(1+ q)^2} \ee where \bea && P_2(t) = 10 t^{+2} + 15 + 10 t^{-2} \nn\\
&& P_3(t) = 20 t^{+3} + 10 t^{+2} + 64 t^{+1} + 22 + 64 t^{-1} +
10 t^{-2} + 20 t^{-3}\nn\\&& P_4(t) = 15 t^{+4} + 8 t^{+3} + 104
t^{+2} + 48 t^{+1} + 175 + 48 t^{-1} + 104 t^{-2} + 8 t^{-3}+ 15 t^{-4}\nn\\
&& P_5(t) = 4 t^{+5} + 2 t^{+4} + 64 t^{+3} + 40 t^{+2} + 196
t^{+1} + 88 + \nn\\ &&\qquad + 196 t^{-1} + 40 t^{-2} + 64 t^{-3}+
2 t^{-4} + 4 t^{-5} \nn\\\eea

Depending on the choice of $k$ one can recognize the surviving 1/2
BPS states as those with $Q=kn$. In formulae one has to replace
$t$ with $\omega^r$ and sum over $r=0, ..., k-1$.

\section{Conclusions}

We have re-analyzed the KK spectrum of $d=11$ supergravity on
$S^7$ and $S^7/\mZ_k$. The latter includes monopole operators dual
to charged states in Type IIA on $\mC\mP^3$. To this end we have
presented some group theoretic methods for the decomposition of
the $SO(8)$ into $SO(6)\times SO(2)$ valid also for other
cosets\cite{Caviezel:2008ik, Haack:2009jg, Kounnas:2007dd} where resolution of the mixing among
various fluctuations should be possible on the basis of symmetry
arguments. In particular, massless vectors associated to Killing
vectors in generic flux vacua with isometries have been recently
discussed in \cite{Bianchi:2010cy}.

We have then considered higher spin symmetry enhancement. We have
displayed the partition functions for singletons, doubletons and
tripletons and discussed in details higher spin fields and 1/2 BPS
states corresponding to KK excitations of $\cN =6$ gauged
supergravity. It would be worth pursuing the analysis to higher
n-pletons and to cases with lower supersymmetry, yet based on
internal coset manifolds.

 \vskip 1cm

 \section*{Acknowledgments}
Discussions with S.~Ferrara, D.~Forcella, F.~Fucito, L.~Lopez,
J.~F.~Morales, R.~Richter, A.~V.~Santini, D.~Sorokin and L.~Wulff
are kindly acknowledged.
M.~B. and M.~S. would like to particularly thank M.~Naghdi for
sharing his insights on the ABJM model and collaborating on
related issues at an early stage of this project. This work was
partially supported by the ERC Advanced Grant n.226455 {\it
``Superfields''} and by the Italian MIUR-PRIN contract 2007-5ATT78
{\it ``Symmetries of the Universe and of the Fundamental
Interactions''}. The work of R.P. was partially supported by the
European Commission FP7 Programme Marie Curie Grant Agreement
PIIF-GA-2008-221571, and the Institutional Partnership grant of
the Humboldt Foundation of Germany.

\section*{Appendix: Dimension formulae for $SO(8)$}

General formula
 \bea
d_{(\ell _1 , \ell _2, \ell _3, \ell
_4)}^{SO(8)}=&&\frac{1}{4320}\times
(1+\ell _1) (1+\ell _2) (1+\ell _3) (1+\ell _4)\nn\\
&& (2+\ell _1+\ell _2) (2+\ell _2+\ell _3) (2+\ell _2+\ell _4) \nn\\
&& (3+\ell _1+\ell _2+\ell _3) (3+\ell _1+\ell _2+\ell _4) (3+\ell_2+\ell _3+\ell _4)\nn\\
&& (4+\ell _1+\ell _2+\ell _3+\ell _4) (5+\ell_1+2 \ell _2+\ell _3+\ell _4) \eea

Specific cases (KK harmonics)

\bea && d_{(\ell , 0, 0, 0)}^{SO(8)}= \frac{1}{360} (1+\ell)
(2+\ell) (3+\ell)^2 (4+\ell) (5+\ell)
\quad \leftrightarrow \quad Y_{N_1}\nn\\
&& d_{(\ell, 1, 0, 0)}^{SO(8)}= \frac{1}{60} (1+\ell) (3+\ell)
(4+\ell)^2 (5+\ell) (7+\ell)
\quad \leftrightarrow \quad Y_{N_7}\nn\\ %
&& d_{(\ell, 0, 1, 1)}^{SO(8)}= \frac{1}{24} (1+\ell) (2+\ell)
(4+\ell)^2
(6+\ell) (7+\ell) \quad \leftrightarrow \quad Y_{N_{21}}\nn\\
&& d_{(\ell, 2, 0, 0)}^{SO(8)}= \frac{1}{18} (1+\ell) (4+\ell)
(5+\ell)^2 (6+\ell) (9+\ell) \quad \leftrightarrow \quad
Y_{N_{27}}\\
&& d_{(\ell, 0, 2, 0)}^{SO(8)}= d_{(\ell, 0, 0, 2)}^{SO(8)}=
\frac{1}{36} (1+\ell) (2+\ell) (3+\ell)
(5+\ell) (6+\ell) (7+\ell) \quad \leftrightarrow \quad Y_{N_{35}}\nn\\
&& d_{(\ell, 0, 1, 0)}^{SO(8)}= d_{(\ell, 0, 0,
1)}^{SO(8)}=\frac{1}{90} (1+\ell) (2+\ell) (3+\ell) (4+\ell)
(5+\ell) (6+\ell)\nn\\
&& d_{(\ell, 1, 1, 0)}^{SO(8)}= d_{(\ell, 1, 0, 1)}^{SO(8)}=
\frac{1}{18} (1+\ell) (3+\ell) (4+\ell) (5+\ell) (6+\ell) (8+\ell)\nn\eea

\section*{Appendix: Zero Charge states}

In this Appendix we list states with $Q=0$ in the KK towers of $S^7$ after the
decomposition of $SO(8)$ into $SO(6)\times SO(2)$.

Bosons:
\bea (\ell,0,0,0)_{\ell\geq 0}\rightarrow
\left[0,\frac{\ell}{2},\frac{\ell}{2}\right]\eea
\bea
(\ell,1,0,0)_{\ell\geq 0}\rightarrow
\left[0,\frac{\ell}{2},\frac{\ell}{2}\right]+\left[0,\frac{\ell}{2}+1,\frac{\ell}{2}+1\right]\nonumber\\
+\left[1,\frac{\ell}{2}+1,\frac{\ell}{2}-1\right
]+\left[1,\frac{\ell}{2}-1,\frac{\ell}{2}+1\right ]
\eea
\bea
(\ell-1,0,1,1)_{\ell\geq 1}\rightarrow
\left[0,\frac{\ell}{2}+2,\frac{\ell}{2}-2\right
]+\left[0,\frac{\ell}{2}-2,
\frac{\ell}{2}+2\right]+\left[1,\frac{\ell}{2}+1,\frac{\ell}{2}-1\right
]\nonumber\\+\left[1,\frac{\ell}{2}-1,\frac{\ell} {2}+1\right
]+\left[0,\frac{\ell}{2},\frac{\ell}
{2}\right]+\left[0,\frac{\ell}{2},\frac{\ell}{2}\right
]\nonumber\\+\left[1,\frac{\ell}{2},\frac{\ell} {2}-2\right
]+\left[1,\frac{\ell}{2}-2,\frac{\ell}{2}\right
]+\left[2,\frac{\ell}{2}-1,\frac{\ell}{2}-1\right] \nn\\
\eea
\bea
(\ell-2,2,0,0)_{\ell\geq 2}\rightarrow
\left[2,\frac{\ell}{2}-3,\frac{\ell}{2}+1\right
]+\left[1,\frac{\ell}{2}-2, \frac{\ell}{2}\right
]+\left[1,\frac{\ell}{2},\frac{\ell}{2}-2\right
]\nonumber\\+\left[0,\frac{\ell}{2},\frac{\ell} {2}\right]
+\left[0,\frac{\ell}{2}-1,\frac{\ell}{2}-1\right]+\left[1,\frac{\ell}{2}-1,\frac{\ell}
{2}+1\right]\nonumber\\+\left[1,\frac{\ell}{2}+1,
\frac{\ell}{2}-1\right]+\left[2,\frac{\ell}{2}-1,\frac{\ell}{2}-1\right]+\left[2,\frac{\ell}{2}+1,
\frac{\ell}{2}-3\right]+\left[0,\frac{\ell}{2}+1,\frac{\ell}{2}+1\right]\nn\\
 \eea
\bea (\ell,0,2,0)_{\ell\geq 0}\rightarrow
\left[0,\frac{\ell}{2}-1,\frac{\ell}{2}+3\right ]+\left[0,
\frac{\ell}{2}+1,\frac{\ell}{2}+1\right ]+\left[0,\frac{\ell}{2}+3,\frac{\ell}{2}-1\right]\nonumber\\
+\left[1,\frac{\ell}{2}-1,\frac{\ell}{2}+1\right ]
+\left[1,\frac{\ell}{2}+1,\frac{\ell}{2}-1\right
]+\left[2,\frac{\ell}{2}-1,\frac{\ell} {2}-1\right]
\eea
\bea
(\ell-2,0,0,2)_{\ell\geq 2}\rightarrow
\left[0,\frac{\ell}{2}-1,\frac{\ell}{2}-1\right]+\left[0,
\frac{\ell}{2}-3,\frac{\ell}{2}+1\right]+\left[1,\frac{\ell}{2},\frac{\ell}{2}-2\right]\nonumber\\
+\left[1,\frac{\ell}{2}-2,\frac{\ell}{2}\right]+\left[2,\frac{\ell}{2}-1,\frac{\ell}{2}-1
\right]+\left[0,\frac{\ell}{2}+1,\frac{\ell}{2}-3\right]
\eea

Fermions:
\bea (\ell,0,0,1)_{\ell\geq 0}\rightarrow
\left[0,\frac{\ell}{2}-1,\frac{\ell}{2}+1\right]+
\left[0,\frac{\ell}{2}+1,\frac{\ell}{2}-1\right]+\left[1,\frac{\ell}{2},\frac{\ell}{2}\right]
 \eea
\bea (\ell-1,0,1,0)_{\ell\geq 1}\rightarrow
\left[0,\frac{\ell}{2}-1,\frac{\ell}{2}+1\right]+
\left[0,\frac{\ell}{2}+1,\frac{\ell}{2}-1\right]+\left[1,\frac{\ell}{2}-1,\frac{\ell}{2}-1\right]\nn\\
\eea
\bea (\ell-1,1,1,0)_{\ell\geq 1}\rightarrow
\left[1,\frac{\ell}{2}-2,\frac{\ell}{2}+2\right]+2\left[1,\frac{\ell}{2},\frac{\ell}{2}\right]
+\left[1,\frac{\ell}{2}+2,\frac{\ell}{2}-2\right]\nonumber\\
+\left[0,\frac{\ell}{2},\frac{\ell}{2}+2\right]+
\left[0,\frac{\ell}{2}+2,\frac{\ell}{2}\right]+
\left[0,\frac{\ell}{2}-1,\frac{\ell}{2}+1\right]\nonumber\\
+\left[0,\frac{\ell}{2}+1,\frac{\ell}{2}-1\right]+
\left[2,\frac{\ell}{2}-2,\frac{\ell}{2}\right]+
\left[2,\frac{\ell}{2},\frac{\ell}{2}-2\right]
+\left[1,\frac{\ell}{2}-1,\frac{\ell}{2}-1\right]
\eea
\bea
(\ell-2,1,0,1)_{\ell\geq 2}\rightarrow
\left[1,\frac{\ell}{2}-3,\frac{\ell}{2}+1\right]
+2\left[1,\frac{\ell}{2}-1,\frac{\ell}{2}-1\right]+
\left[1,\frac{\ell}{2}+1,\frac{\ell}{2}-3\right]\nonumber\\
+\left[0,\frac{\ell}{2}-2,\frac{\ell}{2}\right]+
\left[0,\frac{\ell}{2},\frac{\ell}{2}-2\right]+
\left[0,\frac{\ell}{2}-1,\frac{\ell}{2}+1\right]\nonumber\\
+\left[0,\frac{\ell}{2}+1,\frac{\ell}{2}-1\right]+
\left[2,\frac{\ell}{2}-2,\frac{\ell}{2}\right]+
\left[2,\frac{\ell}{2},\frac{\ell}{2}-2\right]
+\left[1,\frac{\ell}{2},\frac{\ell}{2}\right]\nn\\
 \eea

\section*{Appendix: Generating functions for $SO(8)$ representations}

The generating function for multiplicities of the scalar
spherical harmonics on $S^7$ is given by
\bea
\cF_{N_1}(q)=\frac{1+q}{(1-q)^7}
\eea
The coefficient of $q^\ell$ gives the dimension of the $SO(8)$
representation with Dynkin
label $(\ell,0,0,0)$.

The generating function for vector spherical harmonics with
$SO(8)$ Dynkin label $(\ell-1,1,0,0)$ reads:
\bea
\cF_{N_7}(q)=\frac{(28-36q+35q^2-21q^3+7q^4-q^5) q}{(1-q)^7}
\eea

For two-form spherical harmonics with $SO(8)$ Dynkin label
$(\ell-1,0,1,1)$ the generating function is:
\bea
\cF_{N_{21}}(q)=\frac{(56-42q+22q^2-7q^3+q^4)q^2}{(1-q)^7}
\eea

For second rank symmetric traceless harmonics the $SO(8)$
Dynkin index is $(\ell,2,0,0)$ and the generating function is
given by the following formula:
\bea
\cF_{N_{27}}(q)=\frac{4(75-175q+203q^2-133q^3+47q^4-7q^5)q^2}{(1-q)^7}
\eea

Finally, for three-form spherical harmonics with $SO(8)$ Dynkin
label $(\ell-1,0,2,0)$ (or $(\ell-1,0,0,2)$) one has
\bea
\cF_{N_{35}}(q)=\frac{(35-21q+7q^2-q^3)q^2}{(1-q)^7}
\eea
Let us complete the description with the spectrum of spinor
spherical harmonics.

For gravitini with Dynkin labels $(\ell,0,0,1)_{\ell\geq0}$  and
$(\ell-1,0,1,0)_{\ell\geq1}$, the generating function is:
\bea
\cF_{gravitini}(q)=\frac{8q}{(1-q)^7}
\eea

For spinors with Dynkin labels $(\ell-1,1,1,0)_{\ell\geq1}$ and
$(\ell-2,1,0,1)_{\ell\geq2}$ one has
\bea
\cF_{spinor}(q)=\frac{8q^2(20-35q+35q^2-21q^3+7q^4-q^5)}{(1-q)^7}.
\eea

\section*{Appendix: Generating functions for $SO(6)$ representations}

In this Appendix we present the decomposition of the $SO(8)$
generating functions under $SO(6)\times SO(2)$. Below a factor of
${(1-q t^{-1})^{-4} (1-q t)^{-4}}$ is always understood.

For $(\ell,0,0,0)$ one has: \bea
\hat{\cF}_{graviton}(q)=1-q^2 \eea For $(\ell,1,0,0)$ one
has: \bea
&& \hat{\cF}_{gb1}(q,t)=6 t^2-4 t q-4 t^3 q+q^2+t^4 q^2\nonumber\\
&& \hat{\cF}_{gb2}(q,t)=1-q^2\\
&& \hat{\cF}_{gb3}(q,t)=15+36 q^2- 4 q^3 t^{-3}-4 t^3 q^3+16 q^4+q^6 +(16 q^2+6 q^4)t^{-2}+\nonumber\\
&& t^2 (16 q^2+6 q^4)-(24 q+24 q^3+4 q^5)t^{-1}-t (24 q+24 q^3+4 q^5)\nonumber\\
&& \hat{\cF}_{gb4}(q,t)=6t^{-2}- 4 q t^{-3}- 4 qt^{-1}+q^2+
q^2t^{-4}\nonumber \eea
For $(\ell-1,0,1,1)$ one has: \bea
&& \hat{\cF}_{gb}^1(q,t)=4 t^3 q-6 t^2 q^2-t^4 q^2+4 t q^3-q^4\nonumber\\
&& \hat{\cF}_{gb}^2(q,t)=4 t q-q^2-6 t^2 q^2+4 t^3 q^3-t^4 q^4\nonumber\\
&& \hat{\cF}_{gb}^3(q,t)=-35 q^2+4 t^3 q^3-16 q^4 - 6 q^4t^{-2}-q^6 - t^2(16 q^2+6 q^4)+\nonumber\\
&& (24 q^3+4 q^5)t^{-1}+t (20 q+24 q^3+4 q^5)\nonumber\\
&& \hat{\cF}_{gb}^4(q,t)=-35 q^2 +4 q^3t^{-3}-16 q^4-6 t^2 q^4-q^6 - (16 q^2+6 q^4)t^{-2}+\nonumber\\
&& t (24 q^3+4 q^5) +(20 q+24 q^3+4 q^5)t^{-1}\\
&& \hat{\cF}_{gb}^5(q,t)=4 qt^{-1}-q^2- 6 q^2 t^{-2}+ 4 q^3t^{-3} -q^4t^{-4}\nonumber\\
&& \hat{\cF}_{gb}^6(q,t)= 4 q t^{-3}- q^2t^{-4}- 6 q^2 t^{-2}+ 4 q^3 t^{-1}-q^4\nonumber\\
&& \hat{\cF}_{gb}^7(q,t)=6 t^2 q^2-4 t q^3-4 t^3 q^3+q^4+t^4 q^4\nonumber\\
&& \hat{\cF}_{gb}^8(q,t)= 6 q^4 t^{-2}+6 t^2 q^4- q^2 (20 q+ 4
q^3) t^{-1} - t q^2 (20 q+ 4 q^3) +q^2(20+15
q^2+q^4)\nonumber\\
&& \hat{\cF}_{gb}^9(q,t)= 6 q^2 t^{-2}- 4 q^3 t^{-3}- 4 q^3
t^{-1}+q^4+ q^4 t^{-4}\nonumber \eea
For $(\ell-2,2,0,0)$' one
has: \bea
&& \hat{\cF}_{sc1}^1(q,t)=6 t^2 q^4+6 t^6 q^4 -t^3 q^2 (20 q+4 q^3) -t^5 q^2 (20 q+4 q^3)+t^4 q^2 (20+15 q^2+q^4)\nonumber\\
&& \hat{\cF}_{sc1}^2(q,t)=6 t^2 q^2-4 t q^3-4 t^3 q^3+q^4+t^4 q^4\nonumber\\
&& \hat{\cF}_{sc1}^3(q,t)= 6 q^2 t^{-2}- 4 q^3 t^{-3}- 4 q^3 t^{-1} +q^4 + q^4 t^{-4}\nonumber\\
&& \hat{\cF}_{sc1}^4(q,t)=-1+ 4 q t^{-1}+4 t q-q^2- 6 q^2 t^{-2}-6
t^2 q^2+ 4 q^3 t^{-3}+4 t^3 q^3- q^4 t^{-4}-t^4
q^4\nonumber\\
&& \hat{\cF}_{sc1}^5(q,t)=q^2-q^4\nonumber\\
&& \hat{\cF}_{sc1}^6(q,t)=-36 q^2-16 q^4- 6 q^4 t^{-2}-6 t^6 q^4-q^6-t^2 (6+32 q^2+12 q^4)+ t (24 q+ 28 q^3+ 4 q^5)+\nonumber\\
&& t^3 (24 q+ 28 q^3+ 4 q^5)+ (24 q^3 +4 q^5) t^{-1}+ t^5 (24 q^3+ 4 q^5)-t^4 (36 q^2+16 q^4+q^6)\nonumber\\
&& \hat{\cF}_{sc1}^7(q,t)=-36 q^2-16 q^4- 6 q^4 t^{-6}-6 t^2 q^4-q^6- (6+32 q^2+12 q^4)t^{-2}+ (24 q^3+4 q^5)t^{-5}+\nonumber\\
&& t (24 q^3+4 q^5)+ (24 q+28 q^3+4 q^5) t^{-3}+ (24 q+28 q^3+4 q^5)t^{-1}- (36 q^2+16 q^4+q^6)t^{-4}\nonumber\\
&& \hat{\cF}_{sc1}^8(q,t)= 6 q^4 t^{-2}+6 t^2 q^4- q^2 (20 q+ 4 q^3) t^{-1}- t q^2 (20 q+ 4 q^3)+q^2(20+15 q^2+q^4)\nonumber\\
&& \hat{\cF}_{sc1}^9(q,t)=6 q^4 t^{-6}+ 6 q^4 t^{-2}- q^2(20 q+ 4
q^3)t^{-5}- q^2 (20 q+4 q^3)t^{-3}+ q^2 (20+15
q^2+q^4)t^{-4}\nonumber\\
&& \hat{\cF}_{sc1}^{10}(q,t)=-15-156 q^2-176 q^4- 15 q^4 t^{-4}-15 t^4 q^4-31 q^6+ (60 q^3+24 q^5) t^{-3}+\nonumber\\
&&t^3(60 q^3+24 q^5)- (90 q^2+ 106 q^4+ 16 q^6) t^{-2}- t^2 (90 q^2+ 106 q^4+ 16 q^6)+\nonumber\\
&& (60 q+184 q^3+84 q^5+4 q^7)t^{-1}+ t (60 q+184 q^3+84 q^5+4 q^7)
\eea
For $(\ell,0,2,0)$ one has:
\bea
&& \hat{\cF}_{sc2}^1(q,t)=-4 t^3 q+15 q^2- 4 q^3 t^{-1}+q^4+ t^2 (10+6 q^2)-t (20 q+4 q^3)\nonumber\\
&& \hat{\cF}_{sc2}^2(q,t)=15+36 q^2- 4 q^3 t^{-3}-4 t^3 q^3+16 q^4+q^6+ (16 q^2+ 6 q^4) t^{-2}+\nonumber\\
&& t^2 (16 q^2+ 6 q^4)- (24 q+24 q^3+4 q^5) t^{-1}-t (24 q+24 q^3+4 q^5)\nonumber\\
&& \hat{\cF}_{sc2}^3(q,t)=- 4 q t^{-3}+15 q^2-4 t q^3+q^4+ (10+ 6 q^2)t^{-2}- (20 q+4 q^3)t^{-1}\nonumber\\
&& \hat{\cF}_{sc2}^4(q,t)=4 t^3 q^3- 6 q^4 t^{-2}- t^2 q (16 q+ 6 q^3)+ q (24 q^2+4 q^4)t^{-1}+\nonumber\\
&& t q (20+24 q^2+4 q^4)- q (35 q+ 16 q^3+ q^5)\\
&& \hat{\cF}_{sc2}^5(q,t)= 4 q^3 t^{-3}-6 t^2 q^4- q (16 q+6 q^3) t^{-2}+ q (20+24 q^2+4 q^4) t^{-1}+\nonumber\\
&&  t q (24 q^2+ 4 q^4)-q (35 q+16 q^3+q^5)\nonumber\\
&& \hat{\cF}_{sc2}^6(q,t)= 6 q^4 t^{-2}+6 t^2 q^4- q^2(20 q+4 q^3)
t^{-1}- t q^2 (20 q+4 q^3) +q^2(20 +15 q^2 +q^4)\nonumber \eea
For $(\ell-2,0,0,2)$ one has: \bea
&& \hat{\cF}_{sc3}^1(q,t)=t^4(q^2 -q^4)\nonumber\\
&& \hat{\cF}_{sc3}^2(q,t)=q^2-q^4\nonumber\\
&& \hat{\cF}_{sc3}^3(q,t)=(q^2-q^4)t^{-4}\nonumber\\
&& \hat{\cF}_{sc3}^4(q,t)=6 t^2 q^2-4 t q^3-4 t^3 q^3+q^4+t^4 q^4\\
&& \hat{\cF}_{sc3}^5(q,t)=6 q^2t^{-2}-4 q^3t^{-3}-4 q^3t^{-1}+q^4+ q^4t^{-4}\nonumber\\
&& \hat{\cF}_{sc3}^6(q,t)=6 q^4t^{-2} +6 t^2 q^4 -q^2 (20 q+4
q^3)t^{-1} -t q^2 (20 q +4q^3) +q^2 (20+15 q^2+q^4)\nonumber \eea
For $(\ell,0,0,1)$ one has: \bea
&& \hat{\cF}_{gr1}^1(q,t)=t^2 (1-q^2)\nonumber\\
&& \hat{\cF}_{gr1}^2(q,t)=t^{-2}(1-q^2)\\
&& \hat{\cF}_{gr1}^3(q,t)=6- 4 q t^{-1}-4 t q+ q^2 t^{-2}+t^2
q^2\nonumber \eea
For $(\ell-1,0,1,0)$ one has: \bea
&& \hat{\cF}_{gr2}^1(q,t)=4 t q-6 q^2-t^2 q^2+ 4 q^3 t^{-1}- q^4 t^{-2}\nonumber\\
&& \hat{\cF}_{gr2}^2(q,t)= 4 q t^{-1}-6 q^2- q^2 t^{-2}+4 t q^3-t^2 q^4\\
&& \hat{\cF}_{gr2}^3(q,t)=6 q^2- 4 q^3 t^{-1}-4 t q^3+ q^4
t^{-2}+t^2 q^4\nonumber \eea
For $(\ell-1,1,1,0)$ one has: \bea
&& \hat{\cF}_{f1}^1(q,t)=4 t^5 q^3-6 q^4 -t^4 q (16 q +6 q^3)+ t q (24 q^2+4 q^4) +\nonumber\\
&& t^3 q(20+24 q^2 +4 q^4) -t^2 q (35 q+16 q^3+q^5)\nonumber\\
&& \hat{\cF}_{f1}^2(q,t)= 4 q^3 t^{-1}-6 t^4 q^4-q (16 q+6 q^3)+t q (20+24 q^2+4 q^4)+\nonumber\\
&& t^3 q (24 q^2+4 q^4)-t^2 q(35 q+16 q^3+q^5)\nonumber\\
&& \hat{\cF}_{f1}^3(q,t)= 4 q^3 t^{-5}-6 q^4- q (16 q+6 q^3)t^{-4}+ q(20+24 q^2+4 q^4)t^{-3}+ \nonumber\\
&& q(24 q^2+4 q^4)t^{-1}-q(35 q+16 q^3+q^5)t^{-2}\nonumber\\
&& \hat{\cF}_{f1}^4(q,t)=- 15 q^4 t^{-2}-10 t^4 q^4+ q(56 q^2+24 q^4) t^{-1}+t^3 q(40 q^2+20 q^4)-\nonumber\\
&& t^2 q(60 q+80 q^3+15 q^5)-q (74 q+90 q^3+16 q^5)+t q (36+120 q^2+60 q^4+4 q^6)\nonumber\\
&& \hat{\cF}_{f1}^5(q,t)=4 t q^3- 6 q^4 t^{-4} -q(16 q+6 q^3)+ q (24 q^2+4 q^4) t^{-3}+\nonumber\\
&& q (20+24 q^2+4 q^4) t^{-1}- q (35 q+16 q^3+q^5) t^{-2}\nonumber\\
&& \hat{\cF}_{f1}^6(q,t)=- 10 q^4 t^{-4}-15 t^2 q^4 +t q(56 q^2+24 q^4)+ q (40 q^2+20 q^4) t^{-3}-\nonumber\\
&&  q(60 q+80 q^3+15 q^5) t^{-2} - q(74 q+90 q^3+16 q^5)+ q(36+120 q^2+60 q^4+4 q^6) t^{-1}\nonumber\\
&& \hat{\cF}_{f1}^7(q,t)=4 t q-6 q^2-t^2 q^2+ 4 q^3 t^{-1}- q^4 t^{-2}\nonumber\\
&& \hat{\cF}_{f1}^8(q,t)= 4 q t^{-1}-6 q^2- q^2 t^{-2}+4 t q^3-t^2 q^4\nonumber\\
&& \hat{\cF}_{f1}^9(q,t)=6 q^4+6 t^4 q^4 -t q^2 (20 q+4 q^3) -t^3 q^2 (20 q+4 q^3)+t^2 q^2 (20+15 q^2+q^4)\nonumber\\
&& \hat{\cF}_{f1}^{10}(q,t)=6 q^4+ 6 q^4 t^{-4}- q^2 (20 q+4 q^3)
t^{-3}- q^2 (20 q+4 q^3) t^{-1}+ q^2 (20+15
q^2+q^4) t^{-2}\nonumber\\
&& \hat{\cF}_{f1}^{11}(q,t)=6 q^2- 4 q^3 t^{-1}-4 t q^3+ q^4
t^{-2}+t^2 q^4 \eea
Finally, for $(\ell-2,1,0,1)$ one
has: \bea
&& \hat{\cF}_{f2}^1(q,t)=6 t^4 q^2-4 t^3 q^3-4 t^5 q^3+t^2 q^4+t^6 q^4\nonumber\\
&& \hat{\cF}_{f2}^2(q,t)=6 q^2- 4 q^3 t^{-1}-4 t q^3+ q^4 t^{-2}+t^2 q^4\nonumber\\
&& \hat{\cF}_{f2}^3(q,t)=6 q^2 t^{-4}- 4 q^3 t^{-5}- 4 q^3 t^{-3}+ q^4 t^{-6}+ q^4 t^{-2}\nonumber\\
&& \hat{\cF}_{f2}^4(q,t)=t^2 (q^2-q^4)\nonumber\\
&& \hat{\cF}_{f2}^5(q,t)=(q^2-q^4)t^{-2}\nonumber\\
&& \hat{\cF}_{f2}^6(q,t)=4 t q+4 t^3 q-6 q^2-6 t^4 q^2+ 4 q^3 t^{-1}+4 t^5 q^3- q^4 t^{-2}-t^6 q^4-t^2 (1+q^2)\nonumber\\
&& \hat{\cF}_{f2}^7(q,t)=4 q t^{-3}+ 4 q t^{-1}-6 q^2- 6 q^2
t^{-4}+ 4 q^3 t^{-5}+ 4 t q^3- q^4 t^{-6}-t^2 q^4- (1+
q^2)t^{-2}\nonumber\\
&& \hat{\cF}_{f2}^8(q,t)=6 q^4+6 t^4 q^4 -t q^2(20 q+4 q^3) -t^3
q^2 (20 q+4 q^3)+t^2 q^2 (20+15 q^2+q^4)\nonumber \eea \bea &&
\hat{\cF}_{f2}^9(q,t)=6 q^4+ 6 q^4 t^{-4}- q^2(20 q+4 q^3) t^{-3}-
q^2(20 q+4 q^3) t^{-1}+ q^2(20+15 q^2+q^4)
t^{-2}\nonumber\\
&& \hat{\cF}_{f2}^{10}(q,t)=-6-32 q^2-12 q^4- 6 q^4 t^{-4}-6 t^4 q^4+ (24 q^3+4 q^5)t^{-3}+ t^3 (24 q^3+4 q^5)+\nonumber\\
&& (24 q+28 q^3+4 q^5)t^{-1} +t(24 q+28 q^3+4 q^5) - (36 q^2+16 q^4+q^6) t^{-2} -t^2(36 q^2+16 q^4+q^6)\nonumber\\
&& \hat{\cF}_{f2}^{11}(q,t)=6 q^2- 4 q^3 t^{-1} - 4 t q^3+ q^4
t^{-2}+t^2 q^4 \eea

\end{document}